\documentclass[10pt,preprint]{aastex}
\usepackage{longtable,lscape}

\slugcomment{Version \today}

\begin{document}
\bibliographystyle{apj}
\title {\begin{flushright}{\rm UCRL-JRNL-423904}\\[9mm]\end{flushright}
Reaction rate sensitivity of $^{44}$Ti production in massive stars and implications of a thick target yield measurement of
$^{40}$Ca($\alpha$,$\gamma$)$^{44}$Ti}

\author{
R. D. Hoffman\altaffilmark{1}, 
S. A. Sheets\altaffilmark{1}, 
J. T. Burke\altaffilmark{1}, 
N. D. Scielzo\altaffilmark{1},
T. Rauscher\altaffilmark{4},
E. B. Norman\altaffilmark{1,}\altaffilmark{2,}\altaffilmark{3}, 
S. Tumey\altaffilmark{1},
T. A. Brown\altaffilmark{1},
P. G. Grant\altaffilmark{1},
A. M. Hurst\altaffilmark{1},
L. Phair\altaffilmark{2}, 
M. A. Stoyer\altaffilmark{1},
T. Wooddy\altaffilmark{1}, 
J. L. Fisker\altaffilmark{1},
D. Bleuel\altaffilmark{1}}

\altaffiltext{1}{Lawrence Livermore National Laboratory, 7000 East
  Avenue, Livermore, CA 94551, USA}
\altaffiltext{2}{Nuclear Science Division, Lawrence Berkeley National Laboratory, Berkeley, CA 94720, USA}
\altaffiltext{3}{Nuclear Engineering Department, University of California, Berkeley, CA 94720, USA}
\altaffiltext{4}{Department of Physics, Basel University, Basel, Switzerland}

\begin{abstract}
We evaluate two dominant nuclear reaction rates and their uncertainties that affect $^{44}$Ti production 
in explosive nucleosynthesis.
Experimentally we develop thick-target yields for the $^{40}$Ca($\alpha$,$\gamma$)$^{44}$Ti reaction at
$E_{\rm \alpha} = 4.13, 4.54$, and $5.36$ MeV using $\gamma$-ray spectroscopy. At the highest beam energy, 
we also performed an activation measurement which agrees with the thick target result. From the measured 
yields a stellar reaction rate was developed that is smaller than current statistical-model calculations and recent experimental results,
which would suggest lower $^{44}$Ti production in scenarios for the $\alpha-$rich freeze out. 
Special attention has been paid to assessing realistic uncertainties of stellar reaction rates produced from a combination of experimental and
theoretical cross sections. With such methods, we also develop a re-evaluation of the $^{44}$Ti($\alpha$,$p$)$^{47}$V reaction rate.
Using these two rates we carry out a sensitivity survey of $^{44}$Ti synthesis in eight expansions representing peak temperature and density conditions 
drawn from a suite of recent supernova explosion models. Our results suggest that the current uncertainty in these two reaction rates 
could lead to as large an uncertainty in $^{44}$Ti synthesis as that produced by different treatments of stellar physics.
\end{abstract}


\section{INTRODUCTION}

The dynamic synergy between observation, theory, and experiment developed over many years around the field of $\gamma$-ray astronomy
has as its ultimate goal observations of specifc radionuclides informing our understanding of stellar explosions and the theoretical
models that predict nucleosynthesis. Of the radioactive
species observed so far, the most long lived, $^{26}$Al and $^{60}$Fe
(with half-lives $\tau_{1/2}=7.17\pm 0.24 \times 10^5$ and $2.62\pm 0.04 \times 10^6$ yr, respectively),
are in reasonably good agreement with theoretical predictions \citep{timmes96,diehl06}.
Observations of those in the iron group, $^{56,57}$Ni ($\tau_{1/2}=6.075 \pm 0.01$ d and $35.6 \pm 0.06$ hr, respectively)
and their decay products $^{56,57}$Co ($\tau_{1/2}=77.233\pm 0.027$ d and $271.74\pm 0.06$d, respectively),
are used in many ways to constrain our current models of
the core collapse explosion mechanism.
The radionuclide $^{44}$Ti ($\tau_{1/2}=58.9\pm 0.3$ yr),
made in the same explosive environment but in much lower amounts compared to the very abundant nickle isotopes \citep{wac73,woosley95},
is hoped to one day serve as an even more sensitive diagnostic and a valuable probe of the conditions extant in some of the
deepest layers to be ejected.

Observationally most of the attention 
has focused on the detection of the 68, 78, and 1157 keV $\gamma$-rays from the $^{44}$Ti 
$\rightarrow$ $^{44}$Sc $\rightarrow$ $^{44}$Ca decay chain (see Figure 1). The 1157 keV $\gamma$-ray has been observed directly from a
point source in Cassiopeia A (Cas A; Iyudin et~al. 1994). This was later confirmed by observation of the low-energy 
$^{44}$Sc $\gamma$-rays using the $\textit{BeppoSax}$ \citep{vink01} and $\textit{INTEGRAL}$ \citep{renaud06} observatories. 
Using values for the distance, age, and $\gamma$-flux of Cas A, the amount of $^{44}$Ti ejected was found to be 
1.6$^{+0.6}_{-0.3}$ $\times$10$^{-4}$ M$_\odot$ \citep{renaud06}, in agreement with earlier observations using CGRO \citep{timmes96}. 
Although the presence of $^{44}$Ti is currently below detection limits in SN1987A in the nearby Large Magellanic Cloud, 
its light curve is theorized to now be powered by the decay of $^{44}$Ti. The yield of $^{44}$Ti in SN1987A has been 
estimated from its light curve to be $1 - 2 \times 10^{-4} M_\odot$, a factor of 3 greater than predicted by models \citep{diehl06}.
A third $^{44}$Ti source (of lower significance compared to the one in CasA) 
has been observed in the Vela region \citep{iyudin98} but the existence of a co-located young supernova (SN)
remnant has not been confirmed. While the mass of $^{44}$Ti observed in SN remnants appears to be 
underproduced by past models, the number of observed sources of $^{44}$Ti in all-sky surveys appears to be less than expected from 
estimates of the Galactic SN rate and the known $^{44}$Ti half life, leading some to 
question whether $^{44}$Ti-producing SNe are exceptional \citep{the06}.

Theoretically, $^{44}$Ti production is traditionally ascribed to regions experiencing a strong "$\alpha-$rich freeze out", 
where material initially in nuclear statistical equilibrium (NSE) at relatively low density is cooled so rapidly that 
free $\alpha-$particles do not have time to reassemble, through inefficient $3-$body reactions that span the mass gaps at A=5 and A=8,
back into the iron group. A likely scenario is material in or near the silicon shell as it experiences shock wave
passage during a core collapse SN event. Assuming the material cools adiabatically over a 
hydrodynamic timescale ($\tau_{\rm HD} = 446\chi/\sqrt{\rho_i}$), where $\chi$ is a scaling parameter (here unity) 
and $\rho_i$ is the initial (peak)
density in g cm$^{-3}$, initial conditions that would result in a final $\alpha$-particle mass fraction of $\sim$ 1\%
would require temperatures high enough to ensure NSE ($T_{9i}\gtrapprox 5$)
and an initial density $\rho_i < {\rm min}(4.5\times 10^5 T_{9i}^3, \quad 2.5\times 10^5 T_{9i}^4 \chi^{-2/3})$ \citep{wac73}. 
This translates to a radiation entropy greater than unity (Section 3.2).

In models of massive stars the $\alpha-$rich freeze out dominates the solar production of several species, including
$^{44}$Ca (made as Ti), $^{45}$Sc, $^{57}$Fe (made as Ni), and $^{58,60}$Ni, while still others seem to require
a sizable component to account for their solar abundances, including $^{50,52}$Cr, $^{59}$Co, $^{62}$Ni, and $^{64}$Zn
\citep{woosley95,tnh96}. Although $^{56}$Ni is the dominant species produced in both NSE and the $\alpha-$rich freez out 
over a wide range ($0 \leq \eta \leq 0.01$) of neutron excess \citep{hwei85},
the solar abundance of $^{56}$Fe is dominated by production in explosive silicon
burning in massive stars and by a large contribution from SNe Ia \citep{twoosley95}. 

In order to address one aspect of the model uncertainties associated with the theoretical predictions of $^{44}$Ti in SNe, 
we focus on exploring the nuclear data uncertainties of two key reaction rates.
Experimentally we address the $^{40}$Ca($\alpha$,$\gamma$)$^{44}$Ti cross section where
the existing experimental results are inconsistent and theoretical estimates are complicated by the suppression of 
$E1$ $T = 0 \rightarrow T = 0$ gamma transitions in self-conjugate ($N$ = $Z$) nuclei \citep{rauscher00a}. 
This cross section has been measured in the past using several techniques. 
\cite{cooperman} used in-beam $\gamma$-ray spectroscopy to measure the capture of $\alpha$-particles on a metallic calcium target in the center-of-mass 
energy range $E_{\rm CM} = 2.5 - 3.65$ MeV. In that work the excitation function for the 1083 keV first excited state transition in $^{44}$Ti 
was determined and the resonance strengths were developed. 
\cite{Nassar} determined the integral cross section in the range 
$E_{\rm CM} = 2.2 - 4.17$ MeV by bombarding a He gas target with a $^{40}$Ca beam and collecting the recoiling 
$^{44}$Ti in a catcher foil. Accelerator mass spectroscopy was then used to determine the ratio of $^{44}$Ti/Ti from the 
known content of Ti in the catcher foil. 
Recently, a slightly broader energy range of $E_{\rm CM} = 2.11 - 4.19$ MeV was explored
using the DRAGON recoil mass spectrometer \citep{vockenhuber07} in inverse kinematics. They developed individual resonances whose sum
was used to determine a reaction rate for $1\le T_9 \le 5.5$. 
Yet despite these often heroic expenditures of time, toil, and treasure, in the temperature range of astrophysical 
interest there still exists a factor of 3 or more difference between the experimentally determined reaction rates.

In this work, we develop the cross section for $^{40}$Ca($\alpha$,$\gamma$)$^{44}$Ti by two separate methods as a check on systematic uncertainties.
First we used in-beam $\gamma$-ray spectroscopy to measure a thick target yield of the $^{40}$Ca($\alpha$,$\gamma$)$^{44}$Ti reaction.
We then determined the number of $^{44}$Ti nuclei produced by counting low-energy 
$\gamma$-rays from the decay of $^{44}$Ti in an irradiated target. 
Special attention was devoted to checking the internal consistency of the measurements and to 
establishing realistic uncertainties in developing the stellar reaction rate from a combination of experimental
and theoretical cross sections. We have made a similar evaluation of the stellar reaction rate for the 
dominant destruction reaction, $^{44}$Ti($\alpha$,$p$)$^{47}$V, based on the original experimental work of \cite{sonz00} and
the theoretical cross section work of \cite{rauscher01}.

Our results are presented in four parts. 
In Section 2, we describe our experimental efforts. 
Section 3 then discusses the development of stellar reaction rates, and an
expose of the past and present experimental and theory efforts for the two rates in question.  
In Section 4, we present nucleosynthesis results for the production of $^{44}$Ti and
$^{56,57,58}$Ni reported in previous surveys of massive star evolution. 
We then carry out a sensitivity survey of $^{44}$Ti to variations in the principle 
production and destruction rates using more recent SN models.
In Section 5 we provide a discussion and conclusions.

\section{Experimental Methods }

The $^{40}$Ca($\alpha$,$\gamma$)$^{44}$Ti reaction was measured at the Lawrence Livermore National Laboratory (LLNL) 
Center for Accelerator Mass Spectroscopy (CAMS) using a 10 MV FN Tandem Van de Graaff accelerator. 
To calibrate the beam energy a silicon detector was placed in a low current beam and the measured energy was 
then compared with spectroscopy grade $\alpha$-sources. The $\alpha$ energies used were the 5.30 MeV from $^{210}$Po, 
the 6.12 MeV from $^{252}$Cf, the 5.49 MeV from $^{241}$Am, and the 4.69 and 4.61 MeV $\alpha$'s from the 
decay of $^{230}$Th. From the calibration, the uncertainty in the $\alpha$-beam energy is $\pm$ 5 keV. 
The thick target yield was measured at $E_{\rm \alpha} = 4.13$, $4.54$, and $5.36$ MeV beam energies. 

For each beam energy a target was manufactured by pressing $^{\rm nat}$CaO powder\footnote{Obtained from Alfa Aesar, MA, USA} 
into a copper holder. The powder had a purity of 99.95$\%$ (metals basis) but contained ppm concentrations 
of C and F. To completely stop the beam, each target had a minimum thickness of at least 1.1 mm.  
The target was mounted on a copper block within the vacuum chamber and tilted at 30$^{\circ}$ with respect to the 
beam. The vacuum chamber contained two windows so that the target could be visually monitored. 
A schematic of the target chamber is given in Figure~\ref{chamber}.

The target chamber was electrically isolated from the rest of the beam line. The target was electrically 
connected to the target chamber allowing the beam current to be directly measured from the chamber. 
The current integration from the target chamber was checked to a precision of better than 1$\%$ using a 
NIST-traceable precision DC current source. An opposing-pair magnet was attached upstream of the target 
chamber to suppress the escape of secondary electrons generated by the beam on the target.  
A summary of the irradiation runs is given in Table~\ref{irradiation}.

\begin{deluxetable}{cccccc}
\tabletypesize{\scriptsize}
\tablecaption{Irradiation Conditions for the $^{\rm nat}$CaO Targets.}
\tablewidth{0pt} 
\tablehead{
\colhead{Irradiation Run}
& \colhead{$E_{\alpha}$ (MeV)}
& \colhead{Irradiation Time (hr)}
& \colhead{Total Charge ($\mu$C})
}
\startdata
 1  & 4.13 & 6 & 45808.6 \\
 2 & 4.54 &  7 & 46115.8\\
 3 & 5.36 &  10 & 25763.2 \\
\enddata
\label{irradiation}
\end{deluxetable}

Two 80$\%$ HPGe detectors were used to measure the prompt $\gamma$-ray yield during irradiation. 
One detector was 11.4 cm away from the target at 30$^{\circ}$ with respect to the beam. 
The location of the second detector was at 15.8 cm and an angle of 99$^{\circ}$. 
The HPGe detector thresholds were set at 125 keV. The average deadtime during a run was 10 \%.

The $\gamma$-ray energy spectra was accumulated in 8192 channels using two Ortec AD413a ADCs. 
Efficiency and energy calibrations of the HPGe detectors were made using $^{60}$Co, $^{22}$Na, 
$^{137}$Cs, $^{54}$Mn, and $^{133}$Ba NIST-traceable sources with activities known to a 
$1\sigma$ uncertainty of 1$\%$. Efficiencies of 0.085$\%$ and 0.162$\%$ were found for the 
detectors at 30$^{\circ}$ and 99$^{\circ}$, respectively.  

\subsection{Analysis of Prompt $\gamma$-Ray Data}

The thick target yield for the $^{40}$Ca($\alpha$,$\gamma$)$^{44}$Ti reaction was deduced from the 
yield of the 2$^+$$\rightarrow$0$^+$ 1083 keV transition which collects most of the strength from 
$^{44}$Ti. The angular distribution of the 1083 keV $\gamma$-rays with respect to the angle between 
the $\gamma$-rays and beam direction is given by $W(\theta) = \sum_l a_l P_l(\cos\theta)$ 
with $l$ = 0, 2, 4. The cross section is proportional to $a_0$ so only this term needs to be 
determined \citep{dyer81}. By placing the detectors at angles for which $P$$_4$($\cos\theta$) is zero, 
30.6$^{\circ}$ and 109.9$^{\circ}$, $a_0$ can be determined from measurements at only two angles. 
The experimental thick target yield of the 1083 keV $\gamma$-ray at angle $\theta$ is given by
\begin{equation}
Y_{1083}(\theta) = \frac{N_c}{N_pL_t A_b \,\varepsilon_{1083}},\
\label{yield_eq}
\end{equation}
where $N_c$ is the number of counts in the 1083 keV photopeak, $N_p$ is the number of $\alpha$ particles 
impinging on the target, $L_t$ is the detector live-time fraction, $A_b$ is the natural abundance of $^{40}$Ca, 
and $\varepsilon_{1083}$ is the efficiency of the HPGe detector at 1083 keV. The region near the 1083 keV 
$\gamma$-ray is shown in Figure~\ref{partialspectra} for $E_{\alpha}$ = 5.36 MeV. The 1083 keV photopeak 
lies on the tail of the 1039 keV $^{70}$Ge doppler shifted $\gamma$-ray excited by fast neutrons primarily 
from the $^{19}$F($\alpha,n$) reaction. The background and 1083 keV peak were fit to an error function 
convoluted with a Gaussian, as in \cite{gete97}. The total experimental yield for the 1083 keV $\gamma$-ray 
was determined by finding $a_0$ from the angular distribution of $Y(\theta)$. The experimental yield of the 
1083 keV $\gamma$-ray determined from these fits are given in Table ~\ref{yieldtab}. In order to convert the 
yield of the 1083 keV $\gamma$-ray into the yield of $^{44}$Ti, one must take into account those transitions 
which bypass the 2$^+\rightarrow$ 0$^{+}$ 1083 keV transition in $^{44}$Ti. The Monte Carlo program DICEBOX 
\citep{dicebox} was used to simulate the $\gamma$-ray cascades from the decay of the compound nucleus $^{44}$Ti 
in order to estimate the number of transitions which bypass  the 2$^+\rightarrow$ 0$^{+}$ transition. These 
simulations suggest $(20\pm 3)$\% of transitions bypass the 2$^+\rightarrow$ 0$^{+}$ transition with the error 
on the simulations taking into account the uncertainty in the nuclear level density, the photon strength function, 
and the capture state. In Figure~\ref{thickyield}, the thick target yield corrected for the missed strength 
to the 1083 keV level is compared to the NON-SMOKER cross section.

\begin{deluxetable}{ccccc}
\tabletypesize{\scriptsize}
\tablecaption{$^{40}$Ca($\alpha$,$\gamma$)$^{44}$Ti Measured and Theoretical Thick Target Yields per $\alpha$ 
Particle on Target. $Y_{1083}$ is the measured yield for the 1083 keV $\gamma$-ray, $Y_{^{44}\rm Ti}$ is the total 
yield for the production of $^{44}$Ti, and $Y_{\rm theory}$ the thick target yield calculated from Equation \ref{theoryyield} 
assuming the theory cross section of \cite{rauscher01}. $Y_{\rm offline}$ is the yield from our activation measurement 
of the target irradiated at 5.36 MeV.
\label{yieldtab}}
\tablewidth{0pt}
\tablehead{
\colhead{$E_{\alpha}$ (MeV)}
& \colhead{ $Y_{1083}$ (10$^{-11}$) }
& \colhead{ $Y_{^{44}\rm Ti}$ (10$^{-11}$) }
& \colhead{$Y_{\rm theory}$  (10$^{-11}$)}
& \colhead{$Y_{\rm offline}$  (10$^{-11}$)}
}
\startdata
 4.13       &   2.11 $\pm$ 0.40         &  2.53 $\pm$ 0.50        &       5.96       &    \\
 4.54       &   5.72 $\pm$ 1.1          &  6.86 $\pm$ 1.4         &       16.4       &     \\
 5.36       &   29.2 $\pm$ 5.7          &  35.0$\pm$ 7.1          &       61.0       &  35.7 $\pm$ 2.5     \\
 \enddata
\end{deluxetable}

The experimental yield can be related to a theoretical cross section $\sigma(E)$ by the equation 
\begin{equation}
Y = \int^{E_{\alpha}}_0 \frac{\sigma(E)}{-dE/dx}dE 
\label{theoryyield}
\end{equation}
where $\sigma$($E$) is the energy-dependent cross section, and $dE/dx$ is the stopping power for $^{\rm nat}$CaO. The $dE/dx$ 
values for $^{\rm nat}$CaO were calculated using the program SRIM \citep{ziegler}. Table ~\ref{yieldtab} gives a 
comparison of the experimental and calculated yields using the NON-SMOKER Hauser-Feshbach $^{40}$Ca($\alpha$,$\gamma$)$^{44}$Ti 
cross section \citep{rauscher01}.  The experimental yields are a factor of 1.7-2.3 times smaller than the 
yield calculated from the theoretical cross sections. 

The uncertainties from detector efficiencies, deadtime, angular distribution corrections, coincidence summing, 
and beam current integration are tabulated in Table~\ref{errors}. The integrated beam current was cross checked 
by using the $^{17}$O($\alpha,\alpha\sp{\prime}\gamma$) and $^{44}$Ca($\alpha,\alpha\sp{\prime}\gamma$) reactions. The thick target 
yield for the $^{17}$O($\alpha,\alpha\sp{\prime}\gamma$) reaction that excites  the first excited state at 870.8 keV 
was measured in \cite{Hsu82} at $E_{\alpha}$ = 5.486 MeV. Using Equation (1), $N_p$ from $^{17}$O($\alpha,\alpha\sp{\prime}\gamma$) 
was determined to be (7.2$\pm$0.86$)\times$10$^{16}$ after correcting for the difference in beam energy between \cite{Hsu82} 
and our work. The thick target yield for the Coulomb excitation of the 1157 keV level in $^{44}$Ca was calculated using 
the program GOSIA \citep{Wu83} and used to infer $N_p$ of (8.2$\pm$1.6)$\times$10$^{16}$. This compares well with our 
beam current integrator which gave a value of 7.2$\times$10$^{16}$ for $N_p$. The efficiency of the detector at 
30$^\circ$ was sensitive to the decay location due to the attenuation of $\gamma$-rays through the copper target holder. 
By varying the position of $\gamma$-ray sources to match the approximate 1 cm$^2$ beam spot size the uncertainty in the 
efficiency was determined. The uncertainty due to coincidence summing of $\gamma$-rays in a single detector was estimated 
from the geometric solid angle spanned by the detectors.

\begin{deluxetable}{cc}
\tabletypesize{\scriptsize}
\tablecaption{Compilation of systematic uncertainties for the online measurement (1$\sigma$).}
\tablewidth{0pt} 
\tablehead{
\colhead{Source of uncertainty}
& \colhead{Uncertainty}
}
\startdata
 Beam integration  & 1\%  \\
 Detector efficiency & 6\%-7\% \\
 Deadtime & 1\% \\
 Angular distribution & 17\%-18\% \\
 Coincidence summing & 1\% \\
\enddata
\label{errors}
\end{deluxetable}

The angular distribution is the largest source of uncertainty. Our detectors did not sit at exactly the zeros of the 
$P_4$ term in the angular distribution which introduces some uncertainty when $a_0$ is determined with only two detectors. 
Furthermore, \cite{Simpson71} measured the angular distribution of $\gamma$-rays following resonance $\alpha$ capture 
at $E_\alpha$ = 4.22, 4.26, and 4.52 MeV and found a strong contribution for the $P_4$ term. The uncertainty on $a_0$ 
was estimated by the following procedure. A value was chosen for the $a_4$ coefficient between $\pm$0.5 and then a fit 
of the angular $\gamma$-ray yield was made. The resulting variation in $a_0$ was between 17\% and 18\% for different values 
of $a_4$. This is a conservative estimate because the many $\gamma$-ray cascade paths and alpha energies involved are 
likely to wash out any nuclear alignment and would cause $a_4$ to be nearly zero.

\subsection{Low Background Counting}

The offline counting of the irradiated target took place at the Low Background Counting facility at LLNL. Only the 
target irradiated at $E_\alpha$ = 5.36 MeV was counted because the activity of the other targets was estimated to be too low.  
A HPGe low-energy photon spectrometer (LEPS) detector was used to detect 68 and 78 keV $\gamma$-rays from $^{44}$Ti decay. 
The target was placed 2 mm away from the detector face and counted for two weeks. The activity of the $^{44}$Ti target 
was determined by comparing it to a  56.6 $\pm$ 1.6 nCi $^{44}$Ti reference source. The reference source was counted 
with the LEPS detector in the same geometry as the $^{44}$Ti target. 

The $^{44}$Ti reference source strength was determined by using an 80$\%$ HPGe detector to compare the 1157 keV 
$\gamma$-ray to a $^{22}$Na and $^{60}$Co calibrated source. The count was made at a distance of 60 cm away from 
the detector in order to avoid summing of the 1157 keV $\gamma$-ray with the 511 keV $\gamma$-ray. 

The yield was found from
\begin{equation}
Y = \frac{A\,T_{\frac{1}{2}}}{N_p \ln2}, 
\end{equation}
where $A$ is the activity of the irradiated target and $\tau_{1/2} = 58.9 \pm 0.3$ years is the $^{44}$Ti 
half-life \citep{ahmad06}.

A simultaneous fit of the peaks in Figure~\ref{lepsspectra} gives an experimental yield of (35.7 $\pm$ 2.5) 
$\times$ 10$^{-11}$ $^{44}$Ti per $\alpha$-particle. The uncertainty in the offline yield takes into account 
the uncertainty in the detector efficiency, half-life, integration of beam current, calibration source strength, 
and statistics. Since the activity of the $^{44}$Ti target was determined with a reference source having the same 
decay scheme the need to correct for summing of the 68 and 78 keV $\gamma$-rays was eliminated. This yield is 
22$\%$ higher than the yield from the online counting and agrees with the estimate that (20$\pm$3)$\%$ of 
$\gamma$-ray cascades bypass the 1083 keV level. This also demonstrates that sputtering of the target was 
minimal during particle bombardment. 

\section{Reaction Rates}
\subsection{The True Gamow Window}

Thermonuclear reaction rates are obtained
through integration of the energy-dependent cross section weighted by a Maxwell-Boltzmann (MB) particle distribution \citep{fcz67}. For reaction
$I^{\mu}(j,k)L^{\nu}$
\begin{equation}
\label{eq:reacrate}
N_A \langle {\bar\sigma_{jk}^\mu}v\rangle ={ { 3.732\times 10^{10} }\over{ {\hat A_j}^{1/2}{T_9}^{3/2} } }
\int_0^\infty {\bar\sigma_{jk}^\mu} E_j^\mu \exp(-11.605 E_j^\mu /T_9) dE_j^\mu
\end{equation}
where $N_A$ is Avagadro's number, ${\hat A_j}\sim 3.64$ is the reduced mass of the target and incident projectile in atomic
mass units, $\mu$ is an index representing bound states in the target and product nuclei ($\mu=0$ for the ground state, 1 for first excited, etc.),
and $v$ is the relative velocity of the $\alpha$-particle ($j$) and the target ($I^{\mu}$) in cm s$^{-1}$.
The cross section ${\bar\sigma_{jk}^\mu}$ is expressed in barns, $E_j^\mu$ is the center of mass energy of the $\alpha$-particle and the 
target in MeV, $T_9$ is the temperature in billions of degrees Kelvin, and the reaction rate has units of  cm$^3$ mole$^{-1}$ s$^{-1}$.

There are two important issues regarding the cross section ${\bar\sigma_{jk}^\mu}$ used in the integration and its relation to the
experimental cross section. First, when computing {\it astrophysically relevant} reaction rates, the cross section has to be
the one for a thermally excited target nucleus in the plasma. Only such a stellar cross section allows application of
detailed balance to derive the reverse rate \citep{hwfz76, rauscher09}. Laboratory cross sections are measured with the target nucleus in the ground state
only and the correction, or stellar enhancement factor, has to be calculated from theory. For the $^{40}$Ca($\alpha,\gamma$)$^{44}$Ti and
$^{44}$Ti($\alpha,p$)$^{47}$V reactions the correction factors are unity over the entire temperature range of interest \citep{rauscher00b}.
Therefore the cross section ${\bar\sigma_{jk}^\mu}$ is identical to the cross section
$\sigma$ appearing in Eq.\ \ref{theoryyield}. 

Second, the question arises as to how well the reaction rate is constrained by the experimental cross section data. 
In the rare case of a thick target yield measurement that completely samples the relevant stellar temperature range, a direct calculation
of the thermonuclear reaction rate can be obtained by introducing a simple change of variable and combining Eq.'s \ref{theoryyield} and \ref{eq:reacrate} \citep{roughton76}.
More often limitations of the experimental apparatus and/or beam time allotment restricts the measurement of cross section data to a fairly narrow energy range.
To produce a reaction rate spanning a realistic temperature range for stellar synthesis, one often has to  
supplement the experimental results with theory cross section values to perform the integration in Eq. \ref{eq:reacrate}. The integrand
exhibits a maximum contributing most to the reaction rate due to the folding of the energy-dependences of the cross section and the
MB distribution. Conventionally, the location and width of this Gamow peak are estimated by multiplying a Coulomb
barrier penetration factor with the high-energy tail of the MB distribution or by considering a Gaussian approximation to it
(see, e.g., \cite{claytonbook,rauscher10}). 

However,
this is applicable for capture reactions only as long as the $\gamma$-width is larger than the charged particle width \citep{ilibook}.
Since the $\alpha$ width is changing rapidly with energy due to the Coulomb barrier, this prerequisite may only be
fulfilled at very low projectile energies which effectively shifts the Gamow window to lower energies than expected from the
standard approximation formula. Figure \ref{fig:gamowca40ag} confirms this by showing the ``true'' Gamow window at $T_9=2$ and $T_9=5$
in comparison to the peaks obtained from the standard formula and its approximation for the 
$^{40}$Ca($\alpha,\gamma$)$^{44}$Ti and $^{44}$Ti($\alpha,p$)$^{47}$V reactions. 
Finally, Fig.\ \ref{fig:gamow2exp} compares the energy range spanned by our 
experimental data with the true Gamow window for stellar temperatures $1 \leq T_9 \leq 5$. 
The width of the window is defined such that filling the "window" with experimental data would determine the reaction rate to 10\% accuracy.
The minimum temperature at which an experiment can fully inform the reaction rate is the overlap of the full width of the window and the 
range of experimental data (shaded region). For the astrophysical processes considered
here, the most relevant temperature range is between $2.0 \leq {\rm T}_9 \leq 4.0$ (Section 3.4).
A less severe situation exists for the $^{44}$Ti($\alpha,p$)$^{47}$V (Figure \ref{fig:gamow2expb}).
From these figures it is evident that the limited range of experimental data collected in these two experiments
do not contribute significantly to the rates in the important
temperature range and that the supplemental theory cross sections at lower energies will dominate both
of the rates and their uncertainties. 
Nevertheless, the data can be used to test the NON-SMOKER predictions at
higher energies. 

\subsection{Reaction Rate for $^{40}$Ca($\alpha,\gamma$)$^{44}$Ti}

To obtain a ``semi-experimental'' astrophysical rate for $^{40}$Ca($\alpha,\gamma$)$^{44}$Ti, 
we normalized the NON-SMOKER cross section by a factor of 1.71 to agree
with our measured thick target yield at 5.36 MeV (Figure \ref{thickyield}).
The data at this energy were chosen because the experimental result was confirmed by two independent techniques.
Although the errors on the experimental thick target yield are small,
during the calcuation
of the reaction rate we have to assign an uncertainty which is similar to the uncertainty on the theory cross section.
Therefore we will perform our astrophysical studies
with an assigned factor of 2 uncertainty on this rate. This also accounts for a potentially different energy-dependence of the
theoretical cross section, and fits well with the general factor of $2-3$ accuracy expected for global predictions of low-energy
$\alpha$-capture reactions on self-conjugate nuclei \citep{rauscher00a}.

The resulting reaction rates were then fit assuming the REACLIB parameterization \citep{rauscher01}.
\begin{equation}
\label{eq:7par}
 N_A <\sigma\nu> = \exp(a_0 + a_1{T_9}^{-1} + a_2{T_9}^{-1/3} + a_3{T_9}^{1/3} + a_4{T_9} + a_5{T_9}^{5/3} + a_6\ln({T_9}))
\end{equation}
where $T_9$ is the temperature in $10^9$ K. The fit is defined in the range $0.01 \leq {\rm T}_9 \leq 10.0$.
The reverse rate for $^{44}$Ti($\gamma,\alpha$)$^{40}$Ca is calculated through detailed balance, 
the $Q$ value of the forward reaction is 5.1271 MeV. 

Figure \ref{ca40ag_rates} shows the $^{40}$Ca($\alpha,\gamma$)$^{44}$Ti reaction rates considered in this study. The values are 
generated from fits to reaction rates developed from numerous experimental and theory efforts carried out over the last 20 years.
Some have been used in widely cited nucleosynthesis surveys. In particular, \cite{woosley95} 
utilized the theory rates from \cite{wfhz78}, while \cite{cl06} used the fit published in \cite{rauscher00b}.
\cite{rauscher02} used this same fit but normalized at $T_9=3$ to the "empirical" rate developed in \cite{rauscher00a}.
Only "experimentally" determined rates are shown in the main panel, along with the tabulated Hauser-Feshbach theory based rate (crosses) of
\cite{rauscher01}, hereafter referred to as NON-SMOKER. 
Also plotted in the inset is the ratio of each rate versus the NON-SMOKER rate,
our recommended $^{40}$Ca($\alpha,\gamma$)$^{44}$Ti rate lies between the oldest and most recent experimental results.

For $^{44}$Ti synthesis the most relevant temperature range is $2.0 \leq {\rm T}_9 \leq 4.0$. Over this range
{\sl experimental} $^{40}$Ca($\alpha$,$\gamma$)$^{44}$Ti rates vary by factors of 3-5, respectively. 
The two most recent experimental rates \citep{Nassar,vockenhuber07} are larger than 
our recommended rate and would suggest {\sl increased} $^{44}$Ti synthesis
over our result and the yet smaller original experimental result of \cite{cooperman}. 

The range of variation in the theory rates
is also large, that of \cite{wfhz78} being of the order of the largest experimental results, 
the smallest is clearly that
of \cite{rauscher00b}. Most unfortunately, the rate generated from this published fit differs from the tabulated
NON-SMOKER rate (crosses) by factors of 2 (high and low) on both ends of the important temperature range.
\cite{rauscher00b} fit this rate to an accuracy of 0.92 over the temperature range $0.1 \leq T_9 \le 10$. 
In general fits to charged particle rates have
larger deviations than those for neutron-induced reactions. Most often a quote of low accuracy pertains to
deviation at the lowest temperature points of the fit. Based on arguments related to the nuclear level density and its
impact on the applicability of the statistical model, \cite{rauscher00b} suggest that the lower limit of applicability for this dominant
production rate is $T_9=0.24$, well below the important temperature range in this study.
To be fair the statistical result itself could well be in error by a factor
of 2 or more due to the uncertainty in the global $\alpha+$nucleus optical potential, and
the fit does agree at $T_9=2.8$, which is close
to the midway point of $^{44}$Ti production in several of the scenarios studied in the next section. 
We have re-fit this rate (dotted line in the inset), and observe that 
over the important temperature range it now differs from the NON-SMOKER rate by no more that 12\% (typically better than 5\%). 
We will include both fits in our sensitivity analysis.

With an uncertainty of a factor of 2, our new "semi-experimental" rate encompasses most of the previous experiments and calculations.
This underlines the necessity for further measurements in the relevant astrophysical energy range (2-5 MeV).

\subsection{Reaction Rate for $^{44}$Ti($\alpha,p$)$^{47}$V}

Figure \ \ref{ti44ap_rates} shows a similar comparison for the $^{44}$Ti($\alpha,p$)$^{47}$V rates.
As with the dominant production rate, the original \cite{rauscher00b} fit to this destruction rate
is in poor agreement with the 
statistical NON-SMOKER rate (crosses). They list a fit uncertainty of 0.12 and the lower limit of
applicability as $T_9=0.035$, again outside the important temperature range in this study.
We also note that the NON-SMOKER rate exceeds the older (dot-long-dash line) statistical result of 
\cite{wfhz78} for $T_9\geq 2.4$, which would suggest {\sl decreased} $^{44}$Ti production over that seen in the survey of \cite{woosley95}.

The only published experimental reaction rate \citep{sonz00} is roughly a factor of 2 above the NON-SMOKER rate, which would suggest even
{\sl further decreased}
$^{44}$Ti synthesis compared to the previously available theory rates. 
To generate their reaction rate \cite{sonz00} interpolated their four cross section points
between $E_{\rm c.m.} = 5.7$ and $9.0$ MeV, extended the energy range above and below these values with 
scaled cross sections from SMOKER \citep{tat87},
and then integrated over a MB distribution to obtain the reaction rate and produced a REACLIB fit to it.
They claim agreement with the calculated rate of 43\% at the highest temperatures and 30\% at $T_9=2.5$.
Interestingly, their derived experimental rate shows a very different temperature 
dependence compared to the reaction rates of \cite{rauscher01} and \cite{wfhz78}, suggesting a very different energy dependence of the cross section. 
This result is very surprising as the energy dependence of the SMOKER cross section 
is very similar to that of the NON-SMOKER cross section.
\cite{sonz00} provide plots of their experimental values and calculated cross section between 5 and 10 MeV, 
and the fit of their reaction rate versus SMOKER and \cite{wfhz78}, but unfortunately no further details.
Considering Figures \ref{fig:gamowca40ag} and \ref{fig:gamow2expb} of \cite{sonz00}, it is clear that their reaction rate for $2.5 \leq T_9 \leq 5$ was 
determined completely by the scaled SMOKER cross section values.

We provide a re-evaluation of the rate from the data of \cite{sonz00} which also includes an increased
uncertainty estimate. Since the experimental data (within errors) lies within 20\% of the NON-SMOKER cross section prediction,
we generated a set of hybrid cross sections by supplementing the experimental ones with \textit{unrenormalized} NON-SMOKER
values above and below the experimental energy range. We assigned a factor of 2 uncertainty to the
theory values above the experimental data and a factor of 3 below. For the experimental data points we assigned the errors on energy and cross section
as given in \cite{sonz00}. The code EXP2RATE \citep{exp2rate} allows one to calculate reaction rates from cross sections (or astrophysical
$S$-factors) that include both theoretical uncertainties and experimental error bars. The ratio of the resulting rate with its uncertainty
range to the NON-SMOKER rate is shown in Fig.\ \ref{fig:newsonz}. Again, the experimental data contribute to the rate integral
only at higher temperatures and the standard NON-SMOKER rate is still well within the uncertainty even at those temperatures.
In our sensitivity study we adopt the geometric mean of the upper and lower rate limit as our ``recommended'' rate and consider a variation
of a factor of 3 (up and down) as a reasonable uncertainty estimate based on inspection of the uncertainty limits in the relevant temperature range. 

Note that our semi-experimental rate for $^{44}$Ti($\alpha,p$)$^{47}$V is fit in the endoergic ($-Q$) direction
while the original theory rates \citep{rauscher01,wfhz78} were fit in the exoergic ($+Q$) direction. Reaction rates for targets in 
their ground states alone (or any distribution other than thermal equilibrium) {\it do not obey reciprocity} because
the forward and reverse reactions are not symmetrical \citep{hwfz76}. Thus it is very important 
to measure cross sections for astrophysical application in the direction that is least
affected by excited state effects in the target, which is almost always in the exoergic direction. 
This particular reaction is an exception in that it has a small reaction $Q$ value ($-0.410$ MeV)
and excited state effects that are minimized by Coulomb suppression of the stellar enhancement factor. See \cite{kiss08} 
and \cite{rauscher09} for details.

In Table \ref{ratefits}, we provide fits in the REACLIB format for all of the rates discussed above. 
For nasr06 \citep{Nassar} the reaction rate values we fit were themselves generated from a fit
supplied in that paper, otherwise we have used (and always recommend) tabulated values where available, as we have done
for rath01 \citep{rauscher01}, coop77 \citep{cooperman}, and wfhz78 \citep{wfhz78}. The rates labeled ''hsr10'' are our
evaluated rates from our own experimental work and our re-evaluation of the rate from \cite{sonz00}.
For voch07 \citep{vockenhuber07} we fit their central "Rate" tabulated between $1 \le T_9 \le 5.5$ (see their Table III).
For each fit we define a measure of accuracy, $\zeta$, between the rate $r_i$ and the fit $f_i$
for the seven $T_9$ temperature points 1.0, 1.5, 2.0, 2.5, 3.0, 3.5, and 4.0 as
\begin{equation}
\zeta ={{1}\over{n}}{\sum_{i=1}^{7} \left({{r_i-f_i}\over{f_i}}\right)^{2}}
\end{equation}

We also supply the reverse rate fits in the REACLIB format. {\sl These must be multiplied by the appropriate
ratio of nuclear partition functions} as specified and tabulated in \cite{rauscher00b}. Again, the re-evaluated
and re-fit rates for $^{44}$Ti($\alpha,p$)$^{47}$V must be inserted into REACLIB in the forward rate direction (increasing mass or charge).

\begin{deluxetable}{rrrrrrrrl}
\tabletypesize{\scriptsize}
\tablecaption{Reaction Rate Fit Parameters for $^{40}$Ca($\alpha,\gamma$)$^{44}$Ti and $^{44}$Ti($\alpha,p$)$^{47}$V}
\tablewidth{0pt}
\tablehead{
\colhead{$a_0$}
& \colhead{$a_1$}
& \colhead{$a_2$}
& \colhead{$a_3$}
& \colhead{$a_4$}
& \colhead{$a_5$}
& \colhead{$a_6$}
& \colhead{$\zeta$ }
& \colhead{Author}
}
\startdata
\multicolumn{9}{l}{$^{40}$Ca($\alpha,\gamma$)$^{44}$Ti} \\
  87.90966 &    0.82813 &  -71.88292 &  -38.54585 &    0.23930 &    0.09330 &   12.74610 &  2.71e-04 & nasr06  \\
   0.72761 &  -19.77781 &   -3.31217 &    1.72317 &   -1.92089 &    0.13943 &    7.44775 &  1.43e-04 & voch07   \\
  98.50815 &    1.14177 &  -77.63240 &  -44.20413 &    0.92480 &    0.04292 &   12.16306 &  2.03e-03 & hsr10hi  \\
  97.82161 &    1.14060 &  -77.59144 &  -44.25277 &    0.92716 &    0.04280 &   12.18964 &  2.03e-03 & hsr10rec \\
  97.13505 &    1.13944 &  -77.55049 &  -44.30142 &    0.92953 &    0.04267 &   12.21624 &  2.03e-03 & hsr10lo  \\
-102.45950 &  -34.96400 &   62.07089 &   57.31170 &   -5.70439 &    0.32517 &   -0.02072 &  6.45e-03 & coop77   \\
  85.44962 &    0.67896 &  -73.40626 &  -34.23476 &   -0.04774 &    0.09442 &   10.65082 &  6.63e-03 & rath01  \\
  78.50683 &    0.41907 &  -68.31831 &  -31.11664 &   -0.88984 &    0.16873 &   11.80039 &  8.15e-03 & wfhz78   \\
\multicolumn{9}{l}{$^{44}$Ti($\gamma,\alpha$)$^{40}$Ca}\\
 112.85950 &  -58.66788 &  -71.88292 &  -38.54585 &    0.23930 &    0.09330 &   14.24610 &           & nasr06  \\
  25.67748 &  -79.27381 &   -3.31217 &    1.72317 &   -1.92089 &    0.13943 &    8.94775 &           & vock07   \\
 123.45800 &  -58.35423 &  -77.63240 &  -44.20413 &    0.92480 &    0.04292 &   13.66306 &           & hsr10hi  \\
 122.77150 &  -58.35540 &  -77.59144 &  -44.25277 &    0.92716 &    0.04280 &   13.68964 &           & hsr10rec \\
 122.08490 &  -58.35656 &  -77.55049 &  -44.30142 &    0.92953 &    0.04267 &   13.71624 &           & hsr10lo  \\
 -77.50965 &  -94.46000 &   62.07089 &   57.31170 &   -5.70439 &    0.32517 &    1.47928 &           & coop77   \\
 110.39950 &  -58.81704 &  -73.40626 &  -34.23476 &   -0.04774 &    0.09442 &   12.15082 &           & rath01  \\
\multicolumn{9}{l}{$^{44}$Ti($\alpha,p$)$^{47}$V } \\
 -16.17831 &   -7.55203 &   -3.97859 &    7.78213 &   -3.73270 &    0.26210 &   17.63160 &  1.43e-05 & hsr10hi  \\
 -35.62246 &   -9.04965 &    5.56533 &   18.44151 &   -4.10095 &    0.24244 &   16.05165 &  1.16e-03 & hsr10rec \\
 -55.06717 &  -10.54735 &   15.10957 &   29.10119 &   -4.46916 &    0.22277 &   14.47158 &  4.16e-03 & hsr10lo  \\
  -5.77460 &   -7.03160 &   -9.79294 &    0.64303 &   -2.72164 &    0.19070 &   17.68239 &  8.12e-05 & rath01  \\
\multicolumn{9}{l}{ $^{47}$V($p,\alpha$)$^{44}$Ti } \\
 -14.90159 &   -2.81742 &   -3.97859 &    7.78213 &   -3.73270 &    0.26210 &   17.63160 &           & hsr10hi  \\
 -34.34573 &   -4.31504 &    5.56533 &   18.44151 &   -4.10095 &    0.24244 &   16.05165 &           & hsr10rec \\
 -53.79045 &   -5.81274 &   15.10957 &   29.10119 &   -4.46916 &    0.22277 &   14.47158 &           & hsr10lo  \\
  -4.49788 &   -2.29699 &   -9.79294 &    0.64303 &   -2.72164 &    0.19070 &   17.68239 &           & rath01  \\
\enddata
\label{ratefits}
\end{deluxetable}

Of course many other nuclear reaction rates affect $^{44}$Ti synthesis. A prioritized list was suggested by \cite{the98},
but only for one choice of peak temperature and density ($T_{9p}=5.5, \quad \rho_p = 1.0\times 10^7$ g cm$^{-3}$) and three values 
of electron fraction ($Y_e = 0.5, 0.499$, and 0.497). 
Recently the NuGRID team has presented a preliminary survey of $^{44}$Ti synthesis over a wide range of 
peak temperature and density conditions for $Y_e = 0.5$ in which they define several regions 
where a similar set of reaction rates are suggested to be more effective than others \citep{mag08}. 

For this study we used the reaction rate library 
developed by \cite{rtbl02}. We only explore variations in $^{44}$Ti synthesis due to 
the dominant production and destruction rates mentioned above, but here cite our sources 
for several other key reactions noted by \cite{the98} and \cite{mag08}.

For important rates that produce $^{44}$Ti we adopted the $3\alpha$ reaction rate of
\cite{cf88}. Coupled with their $^{12}$C($\alpha,\gamma$)$^{16}$O rate uniformly multiplied by a factor of 1.7
(corresponding to an $S$-factor at 300 keV of 170 keV barns), these proved optimal for producing the solar abundance
set \citep{twoosley95}. 
Although we do not explore variations in these rates, we acknowledge their importance, especially the former, in setting the 
$\alpha$-abundance on which both $^{40}$Ca and $^{44}$Ti production depend. This choice is not critical to showing the sensitivity
of $^{44}$Ti to the two rates we concentrate on. 
As previously mentioned we developed a new semi-experimental rate for $^{40}$Ca($\alpha$,$\gamma$)$^{44}$Ti
with a factor of 2 uncertainty
and also consider six alternate reaction rates (Figure \ref{ca40ag_rates}).
We also adopted rates for the weakly competing side chain
$^{40}$Ca($\alpha,p$)$^{43}$Sc and $^{43}$Sc($p,\gamma$)$^{44}$Ti from \cite{rauscher00b}. 

The dominant $^{44}$Ti destruction rate in this study was $^{44}$Ti($\alpha,p$)$^{47}$V. We explore three choices, our
re-evaluated \cite{sonz00} rate with uncertainty factors 3 and 1/3 as upper and lower limits. Other destruction
reactions of note were $^{44}$Ti($p,\gamma$)$^{45}$V and $^{45}$V($p,\gamma$)$^{46}$Cr, both taken from \cite{jlf01}.
Finally $^{44}$Ti($\alpha,\gamma$)$^{48}$Cr was taken from \cite{rauscher00b}.

In the following section we adopt our recommended "semi-experimental" rates 
as the default production and destruction rates when we report specific values in our nucleosynthesis survey.

\section{Nucleosynthesis Studies}
\subsection{Initial Conditions}

Previous efforts to explore the sensitivity of $^{44}$Ti synthesis to the $^{40}$Ca($\alpha$,$\gamma$)$^{44}$Ti
reaction rate have often been confined to one-zone parameterized network simulations for 
a limited number of initial conditions \citep{the98, vockenhuber07}. 
A much larger survey promises to explore $^{44}$Ti
synthesis over a wider range of initial conditions \citep{mag08}. From this survey we choose eight points that reflect
peak conditions experienced in various SN explosion models that span
the regimes from incomplete silicon burning to the $\alpha$-rich freeze out (see Table \ref{conditions}).
Ideally one would prefer to explore
rate sensitivity in full star models of stellar evolution \citep{Nassar}, but for species that are made under a limited range of stellar
conditions that do not involve details affected strongly by the stellar physics, such as burning over long time-scales in convective
regions, one zone calculations are usually adequate as long as a reasonable range
of conditions are explored. We take this approach.

Our calculations are straight-forward.
Table \ref{conditions} lists our choices for peak temperature and density. From these we calculate
the hydrodynamic time-scale, $\tau_{\rm HD} = 446\chi/\sqrt{\rho_p}$ and
the radiation entropy,
$S_{\rm rad} = 3.33 \times T_{9p}^3/\rho_{5p}$, where $\rho_{5p} = 10^{-5}\times \rho_{p}$ g cm$^{-3}$ and $S$ is in units
of Boltzmann constant per baryon (hereafter we refer to entropy without its units). For the hydrodynamic timescale $\chi$ is a
scaling parameter (assumed to be unity unless explicitly stated). 
The material then expands adiabatically $(\rho \propto {\rm T}^3)$ with $\rho(t) = \rho_p {\rm exp}(-t/\tau)$ on a
hydrodynamic time-scale until the temperature declines to $T_9 \sim 0.25$, a point where all charged particle reactions
affecting $^{44}$Ti synthesis have frozen out. The last column ($t_{\chi}$) gives the approximate time for each simulation
to terminate based on the scale factor for the hydrodynamic time-scale. We will explore two values, $\chi = 1$ and 5. 
The initial compositions consist of neutron, proton, and $\alpha-$particle mass fractions that provide for a
specific neutron excess $ \eta = \sum_i (N_i - Z_i)(X_i/A_i)$
where $N_i$, $Z_i$, $A_i$ and $X_i$ are the neutron, proton, atomic mass number, and
mass fraction of the isotope $i$ with $\sum_i X_i = 1$.
The neutron excess is related to the electron mole number via $Y_e = 1-2\eta$. 
For each initial condition we will survey a range of neutron excess
$(-0.01 \leq \eta \leq 0.03)$ as well as variations in two specific reaction rates.

\begin{deluxetable}{ccccccccc}
\tabletypesize{\scriptsize}
\tablecaption{Nucleosynthesis Survey - Peak Initial Conditions}
\tablewidth{0pt}
\tablehead{
\colhead{Point}
& \colhead{Model}
& \colhead{$T_{9p}$}
& \colhead{$\rho_{7p}$}
& \colhead{$S_{rad}$}
& \colhead{$\tau_{\rm HD}$}
& \colhead{t$_{\chi=1}$}
& \colhead{$\tau_{\rm HD}\times 5$}
& \colhead{t$_{\chi=5}$}
}
\startdata
  & & $10^9$ K & $10^7$ g cm$^{-3}$ & $k^{-1}$ & sec. & sec. & sec. & sec. \\
1 & CasA   & 6.5 & 0.4  &  22.8 & 0.22  & 2.1  & 1.10 & 10.8 \\
2 & CasA   & 5.5 & 0.2  &  27.7 & 0.32  & 2.9  & 1.60 & 14.3 \\
3 & CasA   & 4.7 & 0.1  &  34.5 & 0.45  & 3.9  & 2.25 & 18.0 \\
4 & 2DExpl & 6.5 & 1.0  &  9.14 & 0.14  & 1.3  & 0.70 &  6.5 \\
5 & 2DExpl & 5.5 & 1.0  &  5.54 & 0.14  & 1.3  & 0.70 &  6.5 \\
6 & 2DExpl & 4.7 & 1.0  &  3.45 & 0.14  & 1.2  & 0.70 &  6.5 \\
7 & 2DMHD  & 6.5 & 10.0 &  0.91 & 0.04  & 0.42 & 0.20 &  2.1 \\
8 & 2DMHD  & 5.5 & 10.0 &  0.55 & 0.04  & 0.41 & 0.20 &  2.1 \\
\enddata
\label{conditions}
\end{deluxetable}

Our peak conditions were chosen to reflect the sensitivity of
the nucleosynthesis to variations in entropy and expansion timescale. 
According to \cite{mag08}, points CasA $1-3$ lie along the track of a model for Cassiopia A and have very similar entropy 
values near 30, they differ from one another by a factor of 1.5 (33\%). 
Points 2DExpl $4-6$ are from a rotating two-dimensional explosion model
(point 5 closely corresponds to a point along the track of a Gamma-ray Burst model), while points 2DMHD-7 and -8 lie along
the track of a two-dimensional rotating MHD star (point 7 samples the "Chasm" region where $^{44}$Ti synthesis is suppressed). We chose
points along these tracks that  exhibit identical peak densities and therefore have constant expansion timescales.
Our aim here is not a full
parameter survey of conditions relevant to $^{44}$Ti synthesis, but rather an exploration of the sensitivity
of $^{44}$Ti synthesis to variations in crucial reaction rates affecting its production (and destruction) under conditions 
relevant to recent models and theory. Lacking detailed composition and velocity structure information
our calculations cannot suggest firm predictions of $^{44}$Ti synthesis for comparison
to observation. Rather this limited survey will take a more general approach
that could serve to place reasonable limits on the production of $^{44}$Ti with respect to $^{56,57,58}$Ni on the basis
of nuclear systematics \citep{Woosley91}.

Our results will be given in terms of normalized
production factors (NPF) $P_{44}$, $P_{57}$, and $P_{58}$, that we define as the production factor for the given
nucleus (the final mass fraction of the species in question divided by
the mass fraction to which it decays in the Sun), normalized to the production factor for $^{56}$Fe. 
Defined in this way NPF's for mass 56 (always made in these expansions predominantly as $^{56}$Ni) are unity. 
Hence any NPF that is greater than
unity will be overproduced with respect to solar iron, any NPF less than unity will be underproduced.
Since roughly one-third to one-half of the iron in the Sun is attributable to SN II \citep{twoosley95}, 
NPF's for radioactive $^{57,58}$Ni up to a factor
of 2 $-$ 3 would be acceptable (i.e. would not violate the solar ratio of $^{57}$Fe/$^{56}$Fe). We consider these as
upper bounds on the synthesis of $^{57,58}$Ni. They also serve to illustrate the range of electron fraction ($Y_e$)
allowed in such expansions. For reference,
$X(^{44}{\rm Ca})_\odot = 1.69\times 10^{-6}$,
$X(^{56}{\rm Fe})_\odot = 1.26\times 10^{-3}$,
$X(^{57}{\rm Fe})_\odot = 2.96\times 10^{-5}$,
$X(^{58}{\rm Ni})_\odot = 5.52\times 10^{-5}$ \citep{lod03}.

The decay of $^{56,57}$Co
(with half-lives of 77.2 and 271.7 days, respectively)
powers the light curves of SN II after the initial hydrogen recombination phase until late times when the longer lived 
$^{44}$Ti and $^{60}$Co (with half-lives of 58.9 and 5.27 years, respectively) are expected to dominate \citep{timmes96}. 
Theory predicts that $^{44}$Ti is always produced in much smaller amounts
than $^{56,57}$Ni, which we also infer from observations of Cassiopia A. 
This sets an upper bound on $P_{44}$. Of course it could always
be less, especially if SN II are not the only source of $^{44}$Ti.

\subsection{$^{44}$Ti Nucleosynthesis in Select Supernova Models}

Before presenting our results we consider production of $^{44}$Ti and $^{56,57,58}$Ni from published models of SN II.
The data are drawn from tables of stellar yields by
\cite{woosley95}, \cite{rauscher02}, \cite{cl06}, and \cite{young06}. 
We calculate mass fractions by normalizing the radioactive yields (in $M_\odot$) to the total mass of material
ejected (initial mass $-$ mass loss $-$ remnant mass) and then form the normalized production factor.

Figure \ref{sn2_npf_sol} 
shows the NPF's for each nucleus versus initial stellar mass for $11 \leq M/M_\odot \leq 25$.
We restrict our consideration to this mass range for two reasons: (1) for these solar metallicity models,
the different treatments of mass loss 
are less pronounced for the lower masses, and (2) each survey explores
a range of explosion energies for models with $M/M_\odot \geq 30$. Both can have a major impact on the ejected mass and the
mass cut at higher masses.
We do however show two results for the 25 $M_\odot$
model of \cite{rauscher02} where the explosion energy was increased to produce a model with double the yield of $^{56}$Ni.
We also present results for a range of explosion energies in the 23 $M_\odot$ model of \cite{young06}.
The yields of \cite{cl06} were derived assuming a constant $^{56}$Ni yield of $0.1 \ M_\odot$.

For $^{57,58}$Ni the results agree with the constraints
imposed by the solar abundances, i.e. production factors between roughly 1 and 3 normalized to the production of $^{56}$Fe in the Sun.
The trends across the mass range shown, especially for $^{57}$Ni, are quite similar, with the exception of the 
20 $M_\sun$ model of \cite{rauscher02} that appears anomalous for reasons well described in that work.

For $^{44}$Ti we see fairly remarkable consistency between the three large surveys 
for the masses between 18 and 25 $M_\odot$ with NPF's ranging between 0.1 and 0.2.
\cite{cl06} never produce it above $0.1$ for any mass, although its uniformity is in
part due to the constant $^{56}$Ni yield ejected. They are however very similar to those of \cite{rauscher02}.
Of note is the larger value for $P_{44}$ in a 23 M$_\odot$ model of Cassiopia A \citep{young06}. The data reflect only their
models with hydrogen envelopes removed to mimic a common-envelope evolution that also assumed a parameterized asymmetry of the
explosion (those without asymetry ejected very little $^{56}$Ni). The error bar reflects
the variation of $P_{44}$ due to their range of simulated explosion energies. A mixing algorithm was also included
(absent in the other three surveys) whose role in the higher production of $^{44}$Ti compared to the other surveys is unclear to us.  

Even at a factor of 2 agreement this uniformity may seem surprising for models that have very different prescriptions for
many important physics elements like mass loss (Woosley \& Weaver had none), convection, opacities, and
reaction rate libraries. \cite{woosley95} was the only one to not use as a base the Hauser-Feshbach reaction rate
compilation of \cite{rauscher00b}. 

More important still for $^{44}$Ti production is the parameterization of the explosion and determination of the mass cut. 
All three of the large surveys use a
parameterization assuming motion of an inner zone (i.e., a piston approach) as opposed to thermal energy input \citep{tnh96}. 
\cite{young06} vary their explosion energies through a parameterization of flux-limited diffusion for neutrinos that
includes a "trapping radius" at which the neutrino opacity and flux is artificially adjusted. It should be noted that
all these efforts initiate the explosion and follow the nuclear burning and subsequent ejecta in one-dimension with the
exception of \cite{young06} who, after 10-100 s, map their results from one-dimension to three-dimensions to follow the mixing
and determine the ultimate ejecta distribution. Unfortunately, their
nucleosynthesis results are preliminary and do not include results for $^{57,58}$Ni, 
which should appear in a forthcoming work. 

\subsection{$^{44}$Ti Nucleosynthesis Sensitivity Survey}

We now consider results for each of our simulated expansions defined in Table \ref{conditions}.
Figure \ref{CasA_npf} shows normalized production factors $P_{44}$, $P_{57}$, and $P_{58}$
versus electron mole number $Y_e$ for adiabatic freeze outs from peak conditions defined for points CasA $1-3$ in Table \ref{conditions}.
All were drawn from a model for Cassiopia A \citep{mag08}.
In the figure, each central point represents a calculation that utilizes our recommended
$^{40}$Ca($\alpha,\gamma$)$^{44}$Ti (production) rate for three choices of $^{44}$Ti($\alpha,p$)$^{47}$V (destruction) rate.
Solid line type and filled squares represent our "recommended" destruction rate, filled triangles represent its upper (dotted) 
and lower (dashed) bound.
The error bars on each central point for all three surveys reflect the minimum and maximum deviations of $P_{44}$ due to the
six other choices of the $^{40}$Ca($\alpha,\gamma$)$^{44}$Ti reaction rate that we considered. 
Tabulated nucleosynthesis results assuming our recommended semi-experimental production and destruction rates are given in Table \ref{nuc_survey}.

The results for CasA points 1 and 2 are very similar. The NPF for $^{57}$Ni 
and solar abundances suggest that the range of allowed electron fraction is $0.4980 \leq Y_e \leq 0.500$.
As expected from nuclear systematics, in this range $^{44}$Ti is always less (by mass fraction) than $^{57,58}$Ni. It is also always 
underproduced with respect to solar iron ($P_{44}$ $\leq$ $P_{56} \equiv 1.0$). An increase in $P_{44}$ is seen at lower $Y_e$, but this is
due to a disparate drop in both $^{56}$Ni and $^{44}$Ti production (for point 2 between $Y_e=0.49$ and 0.485,   
($X^{56}$Ni) dropped by a factor of 46, while ($X^{44}$Ti) dropped by a factor of 3, respectively).  
For our ''recommended'' reaction rates the
normalized production factors are very similar ($\sim 0.1-0.2$ over the entire allowed range of $Y_e$) to those seen in the SN II 
models with masses between 15 and 25 M$_\odot$ (Figure \ref{sn2_npf_sol}). For the lower bound on our destruction rate, they are a factor of 1.7 higher.

For point CasA-3 the results are very different. Now any choice of experimental rate for $^{40}$Ca($\alpha,\gamma$)$^{44}$Ti and
either choice of our ''recommended'' or lower bound for the $^{44}$Ti($\alpha,p$)$^{47}$V destruction rate provides for
nucleosynthesis that makes $^{44}$Ti in proportion to solar iron over the allowed range of $Y_e$. Integrated results from a SN model that 
experiences these conditions in a sizable fraction of its ejecta would be of interest. 
Adopting our upper bound for $^{44}$Ti($\alpha,p$)$^{47}$V still makes $P_{44}$ between 0.2 and 0.8, or up to double the value seen in the two
older stellar model surveys. 
For $Y_e$ less than 0.4980 the normalized production
factors continue to climb but are suggesting a strong overproduction with respect to solar iron ($P_{57} \geq 3$, Table \ref{nuc_survey}).
For $Y_e$ above 0.5, P$_{44}$ is always small, even though the $^{56}$Ni mass fraction remains high ($\sim 0.6$).
We also note that all three of these
expansions were very $\alpha$-rich (the final $\alpha$-mass fractions across the $Y_e$ range illustrated
were 0.20, 0.23, and 0.32 for points CasA 1-3, respectively).

The contour plot of $^{44}$Ti production in
\cite{mag08} suggests declining production for lower peak densities along the track of the CasA model. These translate to
higher entropies (approaching 50 at $T_{9p}=4.2$, $\rho_p = 5.0\times 10^5$ g cm$^{-3}$),
but the temperature is dropping so fast that NSE (needed to make $^{56}$Ni) will be increasingly hard to achieve.
For these peak conditions, assuming $Y_e=0.498$ and our recommended principal production and destruction rates, the final $\alpha$,
$^{44}$Ti, and $^{56,57,58}$Ni mass fractions are $0.45$, $1.8\times 10^{-3}$, 0.38, 0.03, and 0.08 respectively, leading to NPF's $P_{44}=3.53$, 
$P_{57}=3.6$, and $P_{58}=4.8$. This again suggests a strong overproduction of the later two with respect to solar iron. 

For any given expansion the spread in values for $P_{44}$ versus $Y_e$ is straight forward to understand.
In general, the value of $Y_e$ tends to steer the net nuclear flows 
along pathways that are either proton-rich ($Y_e > 0.5$) or neutron-rich ($Y_e \leq 0.5$).
Since $^{44}$Ti and $^{56}$Ni both have $Y_e=0.5$, expansions of material that deviate far from this will
reflect a drop in the abundance of each, although for $^{56}$Ni, which is made in NSE, the change will be less dramatic, especially
for proton-rich conditions \citep{stm08}. For all three points (CasA 1-3) the principle production rate
affecting $^{44}$Ti for all values of $Y_e$ explored is $^{40}$Ca($\alpha,\gamma$)$^{44}$Ti.
Differences in $P_{44}$ across the range of $Y_e$ surveyed are therefore driven principally by the reactions affecting the production
of $^{40}$Ca and the destruction of $^{44}$Ti.

For $0.49 \leq Y_e \leq 0.5$ the dominant sequence of reactions producing $^{40}$Ca are $\alpha$-capture reactions on self-conjugate 
nuclei up to $^{36}$Ar
followed by $^{36}$Ar($\alpha,p$)$^{39}$K($p,\gamma$)$^{40}$Ca, operating principally between $2 \leq {\rm T}_9 \leq 3$.
This is also true for $Y_e > 0.5$, but proton-induced ($p,\gamma$) and ($p,\alpha$) flows provide additional pathways
to build up $^{38}$Ca at the expense of $^{40}$Ca. Proton-capture reactions also compete strongly with
$^{44}$Ti($\alpha,p$)$^{47}$V to accelerate the depletion of $^{44}$Ti over the same temperature range. 
For $Y_e < 0.49$ proton-induced reactions that produce $^{40}$Ca and destroy $^{44}$Ti are suppressed due to 
the lower free proton abundance, the ones affecting production win out. 

As mentioned above, the error bars on each point represent the variation in $P_{44}$ due to the many choices of 
$^{40}$Ca($\alpha,\gamma$)$^{44}$Ti reaction
rate we considered. For point CasA-3 this amounted to roughly 20\% deviations from our ''recommended result'', most often
bracketed by our $\times 2$ upper and lower error bar. Considering only the rates from other efforts, the variation
would have been +20\% and -10\% from our recommended value.
The variation is asymetric due to the relative ratios of the various production rates versus ours which was used to anchor the central point 
(see inset of Figure \ref{ca40ag_rates}). If instead of the factor of 2 errors we assumed for the supplemental theory cross section we
had used the errors suggested by our off-line counting data (Figure \ref{thickyield}), the variation would have been of order 2\%.

The choice of $^{44}$Ti($\alpha,p$)$^{47}$V destruction rate exhibits a larger sensitivity ($\sim 70$\%)
to overall $^{44}$Ti synthesis than the entire spread in the experimental $^{40}$Ca($\alpha,\gamma$)$^{44}$Ti production rates. 
Taken together, the potential uncertainty over the allowed $Y_e$ range in $^{44}$Ti synthesis encompassed by 
the possible choices of production and destruction rate is roughly a factor of 3.

Interestingly, the dot-short-dash lines in Figure \ref{CasA_npf} show $^{44}$Ti synthesis considering the use of
the original published fits \citep{rauscher00b} for both $^{40}$Ca($\alpha,\gamma$)$^{44}$Ti and $^{44}$Ti($\alpha,p$)$^{47}$V, in very close
agreement ($\sim 4\%$) with what resulted using our ''recommended'' rates. 
The reason is that for $1 \leq T_9 \leq 4$, the original fit to the production
rate (Figure \ref{ca40ag_rates}) made {\sl on average} (more above $T_9=2.4$, less below) very nearly the same amount of $^{44}$Ti as our 
''recommended rate'', while the destruction
rate (Figure \ref{ti44ap_rates}) was nearly identical for $T_9\geq 3$. Our higher production rate below $T_9=2.4$ accounted for the uniformly
higher production. Again, this is the same level of variation from our recommended rate that would have occurred if we had considered only our
offline counting error (Table \ref{thickyield}). This would also be the result obtained using the default BDAT reaction rate library provided with the
TORCH reaction network code (Timmes 2010, private communication), 
since it utilizes the same (original) fits to the relevant Hauser-Feshbach theory rates \citep{rauscher00b} 
and $3\alpha$ reaction rate \citep{cf88}.

Figure \ref{2DEx_npf} shows our results for points 2DExpl $4-6$ (Table \ref{nuc_survey}) 
drawn from conditions 
in a rotating two-dimensional SN explosion model. These have the same peak temperature albeit at higher peak density, as the points drawn from
the CasA model, and sample the same regions as depicted in \cite{mag08}.
The results for $P_{57}$, and $P_{58}$ are very similar to those from the model for CasA
(i.e. production compared to solar iron), but $P_{44}$ is much smaller for all scenarios. It is similar however to that seen in
the massive star surveys (P$_{44} \sim 0.1-0.2$, Figure \ref{sn2_npf_sol}).
The final $\alpha$-particle mass fractions across the $Y_e$ range illustrated 
were roughly 0.10, 0.07, and 0.06 for points 4-6, respectively (Table \ref{nuc_survey}). 

According to \cite{mag08} points 2DMHD $7-8$ represent
conditions in a model for a rotating two-dimensional MHD star. Point 7 is located in the "Chasm" of $^{44}$Ti production, while point 8
is in the region defined by them as incomplete silicon burning. Both experience much higher peak densities (at
identical peak temperatures) than points from the previous two models, and consequently they have very low 
entropies and very short hydrodynamic time scales.
Results assuming our ''recommended'' production and destruction rates are shown in Table \ref{nuc_survey}.
As before $P_{57}$ and $P_{58}$ are very similar to our previous results reflecting a near constant
production of $^{57,58}$Ni with respect to solar iron (for a given $Y_e$) over the entire range of expansion timescales surveyed. 
However, $P_{44}$ was virtually non-existent,
with ($X^{44}$Ti) $\leq 10^{-5}$ (often much less) for any $Y_e$ in either expansion.
Considering point 7, whose peak conditions reside in the "Chasm" of $^{44}$Ti production noted in \cite{mag08}, our
''recommended'' reaction rates would slightly lessen (by 4\%) the "depth" of the chasm 
compared to the reaction rates they used.

The explanation for the drop in $P_{44}$ stems from the lower entropy (higher peak density)
for each point in these rotating two-dimensional explosion model expansions versus those from the model for CasA 
(see Table \ref{nuc_survey}). Similarly, for all three CasA points, the lower entropy expansions 
translated into a lower $P_{44}$ (Figure \ref{CasA_npf}).
At lower entropy material merges very quickly into the iron group at the expense of the $\alpha$-particle abundance     
which at late times was so low ($\sim 1$\%) in the two-dimensional explosion model expansions that few were available
to make $^{40}$Ca at appreciable levels, with $^{44}$Ti being consequently even lower.

From the track of the rotating two-dimensional explosion model we chose conditions (points 2DExpl $4-6$) that had identical peak densities, and
hence a constant expansion timescale (0.14 s). 
For this factor of 2.6 overall change in entropy and for the specific 
case of $Y_e=0.4980$, the final $^{56}$Ni mass fractions were 0.74, 0.77, and 0.77 for 
points 2DExpl $4-6$, respectively. For the CasA model, with a factor of 1.5 range in entropy (that included a factor of 2
range of dynamic time-scale) the $^{56}$Ni mass fractions were 0.64, 0.62, and 0.52 for points CasA 1-3, respectively.
The salient point for $^{44}$Ti synthesis again is that the higher entropy expansions made less $^{56}$Ni and had 
more $\alpha$-particles, especially at late times, that enabled the production of $^{40}$Ca and $^{44}$Ti (Table \ref{nuc_survey}). 
Since P$_{44}$ is a ratio of Ti to Fe, it grows as a function of increasing entropy. 

We illustrate all these points graphically in Figure
\ref{3panel_msfvst9} where we show the evolution of select mass fractions versus temperature
for the three expansions
in our survey with identical peak temperatures (points 2, 5, and 8) that all started with an initial composition with $Y_e = 0.4980$. 
For each expansion both $^{40}$Ca and $^{44}$Ti achieve an NSE abundance not very different from their final freeze out values,
but both are effectively destroyed for $4 \leq T_9 \leq 5$. Thereafter they are reassembled at a rate determined largely 
by the $\alpha$-particle abundance, which is affected by the peak density (entropy). It is clear that a more robust
$\alpha$-rich freeze out is more conducive to $^{44}$Ti synthesis and that reaction rates clearly matter.

We also note for all conditions shown in Figure \ref{3panel_msfvst9} the freeze-out of the $^{44}$Ti abundance near $T_9=2$. 
The $^{47}$V abundance does increase as the
temperature declines, but the $^{44}$Ti abundance changes by less than a few percent (the increase is due 
predominantly to the decay of radioactive $^{47}$Cr). At this $Y_e$ and 
for ${\rm T}_9 \geq 2.0$ the dominant destruction rate for $^{44}$Ti
is $^{44}$Ti($\alpha,p$)$^{47}$V, below 1.0 it switches to $^{44}$Ti($p,\gamma$)$^{45}$V.
Terminating our simulations at any point below $T_9=2.0$ would have a negligible effect on $P_{44}$.

\subsubsection{Effect of the Expansion Timescale}

$^{44}$Ti synthesis is also affected by the expansion timescale,
$\tau_{\rm HD} = 446\chi/\sqrt{\rho_p}$. 
Figure \ref{3yep4980psi1ovr} shows the range of nuclei produced
in an $\alpha$-rich freeze-out from peak conditions given for point CasA-3 in Table \ref{conditions}
with an initial composition of $Y_e = 0.4980$ and our default scaling of $\chi = 1$.
The nucleosynthesis is presented in terms of ''traditional'' production factors
(so far our results have been expressed as ratios of production factors to that of $^{56}$Fe).
In the figure isotopes of a given element are connected
by solid lines, those produced as radioactive progenitors are surrounded by a diamond, the most
abundant isotope of a given element is denoted by a star. The dotted lines indicate a factor of
2 above and below the dashed line centered on $^{56}$Fe (made as $^{56}$Ni, $^{44}$Ca is made as $^{44}$Ti).
The dominant species are in the iron group. This is a fairly typical
result for any of the CasA or two-dimensional explosion model expansions within the allowed range of $Y_e$.

Figure \ref{ratioovr15} explores the effect of increasing the hydrodynamic time-scale by a factor of $\chi = 5$.
Shown is a straight forward ratio of ''traditional'' production factors for the CasA-3 expansion assuming
both the default ($\chi = 1$) and extended ($\chi = 5$) scaling on $\tau_{\rm HD}$.
The species in the iron group are not much affected, $^{56,58}$Ni are slightly higher than unity, all others 
show lower production factors. The reason is that for a longer expansion timescale,
the material experiences higher temperatures for a longer duration. Those made dominantly
in NSE (the iron group seen here) are slightly altered
due to $\alpha$-particle reactions on them, which are ultimately provided by photo-disintegration and operation of the
$3\alpha$-reaction. The effect is to reduce the overall fraction of $\alpha$-particles
available at late times for the re-assembly of species lighter than the iron group. 

Figure \ref{xalphapsi5db1} shows for the specific CasA-3 example above the temperature evolution over 
the first 4 s  assuming the two scalings.
Over that time the temperature in the default ($\chi = 1$, dotted line) scenario declined from  
$T_9 = 4.7$ - 0.25 while the longer expansion ($\chi = 5$, dashed line) only declined to $T_9 = 2.6$.
Also shown is the ratio of the $\alpha$-particle mass fraction (long/default). 
Recall that both $^{40}$Ca and $^{44}$Ti begin to re-assemble below $T_9 = 4$ (Figure \ref{3panel_msfvst9}).
By the time the longer expansion scenario reaches $T_9 = 4$, it only has 65\% of the $\alpha$-particle mass fraction
that the default expansion had, by $T_9 = 3$ it is down to 52\%. With only half the 
$\alpha$-particles to work with, the expansion with the higher scaling only reached $P_{44} = 0.5 \pm 0.1$,
a factor of 2.6 less than the default expansion. Figure \ref{CasAnpf5} shows the effect over all scenarios considered
in the CasA model expansions. The average $P_{44}$ for points 4-6 (the two-dimensional explosion model) were 0.1, 0.05, and 0.08,
respectively,
over the allowed range of $Y_e$. For points 7 and 8 they are essentially zero.

\section{Conclusions}

We have considered the sensitivity to $^{44}$Ti production in expansions that approximate freeze outs from NSE
due to variations in the principle 
production and destruction reactions $^{40}$Ca($\alpha,\gamma$)$^{44}$Ti and $^{44}$Ti($\alpha,p$)$^{47}$V, and contrast them 
to experimental and theory reaction rate developments over the past 20 years.
Experimentally we have also measured a thick-target yield for the $^{40}$Ca($\alpha$,$\gamma$)$^{44}$Ti reaction.
In-beam $\gamma$-ray spectroscopy was used to determine the yield of the 1083 keV prompt $\gamma$-ray from 
$^{44}$Ti at $E_\alpha$ = 4.13, 4.54, and 5.36 MeV. In order to correct for those transitions which bypass the 
1083 keV transition, the Monte Carlo code DICEBOX was used to estimate a correction of 20$\%$ to the in-beam thick target yield. 
An off-line activation measurement using the target from the $E_\alpha$ = 5.36 MeV irradiation showed
good agreement with the in-beam measurement. 

We then derived a thermonuclear reaction rate by normalizing the NON-SMOKER cross section (down by a factor of 1.71)
to agree with our measured off-line thick target yield. We derived an error bar for our recommended rate 
whose magnitude ($\pm \times 2$) was dominated by the {\sl theoretical} cross section error.
We also carry out a similar re-evaluation of the $^{44}$Ti($\alpha,p$)$^{47}$V reaction rate whose cross section was measured by \cite{sonz00}.
For both reactions, we conclude that the experimental data were far above the Gamow window, and suggest that further
measurements be attempted.

We then carried out a sensitivity survey of
$^{44}$Ti nucleosynthesis in adiabatic expansions from eight peak temperature and density combinations
drawn from conditions in three recent stellar explosion models \citep{mag08}. For each expansion we survey
a range of initial compositions ($0.505 \geq Y_e \geq 0.485$). We also vary the principle production and destruction 
rates affecting $^{44}$Ti in these expansions (eight choices of production rate, four for destruction). Our results
show $^{44}$Ti produced in proportion to solar iron for only one expansion drawn from a model for Cassiopia A, even
though the final mass fractions for $^{44}$Ti in most of the expansions are typically of order $10^{-4}$.
With one exception, the other expansions are consistent with $^{44}$Ti production seen in previous surveys of one-dimensional
stellar evolution and
with constraints imposed by solar abundances. Our results suggest that a strong $\alpha$-rich freeze out ($X(\alpha)_f\sim 0.2-0.3$) 
is highly conducive to $^{44}$Ti synthesis.

With respect to reaction rate sensitivity, our experimental results suggest a recommended
$^{40}$Ca($\alpha,\gamma$)$^{44}$Ti reaction rate that is smaller than those predicted by the most recent experimental efforts
\citep{vockenhuber07,Nassar}, but with a fairly large error bar ($\pm \times 2$) 
that would in fact encompass the former one. Nucleosynthesis models using 
our recommended rate would suggest less $^{44}$Ti than these two recent experiments,
although it would be higher than that suggested by an earlier measured rate \citep{cooperman} and the current theory rate \citep{rauscher00b}.
The total range of sensitivity is a factor of 1.5 (36\%) when considering all of the production rates available.
We also find that the uncertainties associated with the dominant destruction rate, $^{44}$Ti($\alpha,p$)$^{47}$V, 
have roughly double the impact on $^{44}$Ti synthesis ($\times 3.2$, or 70\%) than that exhibited by the entire range of production rates. We suggest the
use of our re-evaluation of the only available experimental reaction rate \citep{sonz00} and strongly suggest consideration of
its attendant larger uncertainty ($\pm \times 3$).
This is a slightly larger spread in $^{44}$Ti sensitivity than observed in two recent surveys of massive star nucleosynthesis \citep{rauscher02,cl06} that used
essentially the same nuclear data, suggesting that current uncertainties in reaction rates could lead to as large
an uncertainty in $^{44}$Ti synthesis as that produced by different treatments of stellar physics.
 
Since the seminal
work of \cite{wac73}, several new and novel theories of SN nucleosynthesis have featured regimes within the
neutrino wind where the $\alpha-$rich freeze out plays
an important role, including scenarios for the $r$-process \citep{woo94,hwq97}, and the $\nu p$-process
\citep{frolich06,pruet06}. In each a combination of high entropy $(S_{\rm rad}\geq 50)$ and short expansion time-scale are required to produce
the unique signatures of each process (the $r$-abundances, and light $p$-nuclei, respectively).
However, as a consequence of extreme neutrino irradiation, the composition is forced away from
neutron-proton equality and ultimately high entropy material enters regions above the iron group where
local effects due to rapid changes in particle separation energies have
a strong influence on the net nuclear flows. In these works no $^{44}$Ti is reported.

Further out in the SN ejecta theory does not show $^{44}$Ti enhancement due to
the $\nu$-process in massive stars \citep{whhh90,woosley95}, nor in recent low-mass (electron capture) core-collapse scenarios
\citep{hmj08,wan09}. The later has been shown to effectively synthesize the long elusive $\alpha$-rich freeze-out candidate $^{64}$Zn, but not $^{44}$Ti.
Our limited survey suggests that $^{44}$Ti synthesis requires modest entropy $(S_{\rm rad}\sim 35$) and expansion timescales 
($\sim \tau_{\rm HD} = 0.45$ s) over a fairly narrow range of $Y_e$ ($0.5 \geq Y_e \geq 0.4980$).

To date surveys of massive star evolution and nucleosynthesis have typically underproduced species whose production is attributed 
to the $\alpha$-rich freeze-out, in particular $^{44}$Ca (made as $^{44}$Ti) and $^{64}$Zn \citep{woosley95,cl06}. This has largely 
been ascribed to variations in treatments of stellar physics, most notably the parameterization of the explosion in one-dimensional models.
Incorporating stellar yields from these surveys into models for galactic chemical evolution indicate the degree of 
underproduction, roughly a factor of a $2-3$ \citep{twoosley95}.
Recent work has lead to models for the progenitor of Cassiopia A that suggest increased production, but the 
underlying physics is still quite uncertain and in need of improvement \citep{young06}. Future attention may focus on issues
related to {\sl asymmetrical} explosions where a noted increase in $^{44}$Ti production compared to models with
imposed spherical symmetry has been suggested for many years \citep{nag97,nag98,hwang03,young06}. The community is on
the threshold of three-dimensional calculations that should provide valuable insight into the core-collapse mechanism, including physics such as
rotation and magnetic fields which will likely enforce an asymmetrical result. Such results should  
help guide future parameterizations of the explosion in our one-dimensional models which will continue to carry the burden in future surveys of
massive star evolution and nucleosynthesis. We believe we have addressed the uncertainty in two key nuclear reaction rates affecting $^{44}$Ti
synthesis, but ultimately, if Type II SNe are the dominant site of $^{44}$Ti production, future models will have to include
a larger fraction of their ejecta that experience an $\alpha$-rich freeze out than they have in the past. Another solution would be
an additional source of $^{44}$Ti, such as rare SNe of type Ia \citep{wtw86,woo97} or Ib \citep{perets10}. 

\acknowledgments

We thank Irshad Ahmad and John Greene of Argonne National Laboratory for preparing the $^{44}$Ti calibration source.
We also thank Cindy Conrado and Kelly Burke for technical assistance. This work was performed under the auspices of the 
U.S. Department of Energy at Lawrence Livermore National Laboratory under Contract No. DE-AC52-07NA27344 and at the
University of California, Lawrence Berkeley National Laboratory under Contract No. W-7405-Eng-48.
Additional support was granted through the DOE Scientific Discovery through Advanced Computing program (DC-FC02-01ER41176),
and by the Swiss National Science Foundation (grant 2000-105328).

\clearpage


\clearpage

\begin{center}
\begin{longtable}{ccccccccccc}
\multicolumn{11}{c}{Table 6. Nucleosynthesis Survey$^1$} \\
\hline \hline
$Y_e$        &     0.5050 &     0.5000 &     0.4995 &     0.4990 &     0.4985 &     0.4980 &     0.4965 &     0.4950 &     0.4900 &     0.4850  \\
$\eta$       &    -0.0100 &     0.0000 &     0.0010 &     0.0020 &     0.0030 &     0.0040 &     0.0070 &     0.0100 &     0.0200 &     0.0300  \\
\hline
\multicolumn{11}{c}{Point 1: Model for Cassiopia A,  $T_{9p}= 6.5,   \rho_{p}=4.0\times 10^6$,   S$_{\rm rad} = 22.8,    \tau_{\rm HD} = 0.22$ s} \\
$X(\alpha$)  &   2.06(-1) &   2.08(-1) &   2.07(-1) &   2.06(-1) &   2.04(-1) &   2.03(-1) &   1.98(-1) &   1.94(-1) &   1.79(-1) &   1.64(-1) \cr
$X(^{40}$Ca) &   1.66(-4) &   8.27(-4) &   7.87(-4) &   7.46(-4) &   7.15(-4) &   6.89(-4) &   6.26(-4) &   5.72(-4) &   4.25(-4) &   1.00(-4) \cr
$X(^{44}$Ti) &   6.17(-5) &   3.00(-4) &   3.49(-4) &   3.42(-4) &   3.34(-4) &   3.27(-4) &   3.05(-4) &   2.85(-4) &   2.22(-4) &   7.39(-5) \cr
$X(^{56}$Ni) &   7.51(-1) &   7.35(-1) &   7.17(-1) &   6.92(-1) &   6.66(-1) &   6.39(-1) &   5.58(-1) &   4.77(-1) &   2.11(-1) &   4.58(-3) \cr
$X(^{57}$Ni) &   2.09(-3) &   1.22(-2) &   2.89(-2) &   3.09(-2) &   3.27(-2) &   3.42(-2) &   3.78(-2) &   3.98(-2) &   3.48(-2) &   4.52(-3) \cr
$X(^{58}$Ni) &   2.25(-5) &   1.57(-3) &   9.70(-3) &   3.50(-2) &   6.04(-2) &   8.59(-2) &   1.63(-1) &   2.41(-1) &   5.08(-1) &   6.97(-1) \cr
$P_{44}$     &      0.062 &      0.305 &      0.364 &      0.369 &      0.374 &      0.382 &      0.409 &      0.446 &      0.786 &     12.027 \cr
$P_{57}$     &      0.131 &      0.707 &      1.717 &      1.891 &      2.079 &      2.283 &      2.889 &      3.536 &      7.024 &     42.192 \cr
$P_{58}$     &      0.004 &      0.221 &      0.309 &      1.153 &      2.060 &      3.071 &      6.659 &     11.504 &     54.762 &   3452.055 \cr
\hline
\multicolumn{11}{c}{Point 2: Model for Cassiopia A,  $T_{9p}= 5.5,   \rho_{p}=2.0\times 10^6$,   S$_{\rm rad} = 27.7,    \tau_{\rm HD} = 0.32$ s} \\
$X(\alpha$)  &   2.22(-1) &   2.25(-1) &   2.25(-1) &   2.23(-1) &   2.22(-1) &   2.21(-1) &   2.16(-1) &   2.12(-1) &   1.98(-1) &   1.84(-1) \cr
$X(^{40}$Ca) &   2.47(-4) &   9.50(-4) &   9.06(-4) &   8.61(-4) &   8.27(-4) &   7.99(-4) &   7.30(-4) &   6.73(-4) &   5.15(-4) &   1.04(-4) \cr
$X(^{44}$Ti) &   8.66(-5) &   3.59(-4) &   4.14(-4) &   4.06(-4) &   3.98(-4) &   3.90(-4) &   3.68(-4) &   3.46(-4) &   2.80(-4) &   8.13(-5) \cr
$X(^{56}$Ni) &   7.35(-1) &   7.17(-1) &   6.99(-1) &   6.75(-1) &   6.48(-1) &   6.22(-1) &   5.40(-1) &   4.59(-1) &   1.92(-1) &   3.06(-3) \cr
$X(^{57}$Ni) &   2.18(-3) &   1.22(-2) &   2.82(-2) &   3.00(-2) &   3.17(-2) &   3.32(-2) &   3.65(-2) &   3.82(-2) &   3.25(-2) &   3.17(-3) \cr
$X(^{58}$Ni) &   2.53(-5) &   2.03(-3) &   1.00(-2) &   3.54(-2) &   6.09(-2) &   8.64(-2) &   1.64(-1) &   2.42(-1) &   5.09(-1) &   6.57(-1) \cr
$P_{44}$     &      0.088 &      0.374 &      0.441 &      0.450 &      0.458 &      0.468 &      0.508 &      0.562 &      1.085 &     19.836 \cr
$P_{57}$     &      0.132 &      0.721 &      1.715 &      1.884 &      2.078 &      2.267 &      2.867 &      3.534 &      7.190 &     44.672 \cr
$P_{58}$     &      0.004 &      0.216 &      0.328 &      1.198 &      2.136 &      3.178 &      6.900 &     12.000 &     60.261 &   4877.049 \cr
\hline
\multicolumn{11}{c}{Point 3: Model for Cassiopia A,  $T_{9p}= 4.7,   \rho_{p}=1.0\times 10^6$,   S$_{\rm rad} = 34.5,    \tau_{\rm HD} = 0.45$ s} \\
$X(\alpha$)  &   3.13(-1) &   3.17(-1) &   3.16(-1) &   3.15(-1) &   3.14(-1) &   3.13(-1) &   3.09(-1) &   3.06(-1) &   2.96(-1) &   2.86(-1) \cr
$X(^{40}$Ca) &   7.70(-4) &   2.10(-3) &   2.04(-3) &   1.94(-3) &   1.87(-3) &   1.82(-3) &   1.69(-3) &   1.59(-3) &   1.31(-3) &   1.85(-4) \cr
$X(^{44}$Ti) &   2.05(-4) &   8.57(-4) &   9.81(-4) &   9.69(-4) &   9.57(-4) &   9.45(-4) &   9.10(-4) &   8.77(-4) &   7.74(-4) &   1.44(-4) \cr
$X(^{56}$Ni) &   6.40(-1) &   6.17(-1) &   5.99(-1) &   5.75(-1) &   5.49(-1) &   5.22(-1) &   4.41(-1) &   3.60(-1) &   9.43(-2) &   2.56(-3) \cr
$X(^{57}$Ni) &   1.83(-3) &   1.27(-2) &   2.81(-2) &   2.99(-2) &   3.15(-2) &   3.29(-2) &   3.55(-2) &   3.62(-2) &   2.39(-2) &   1.30(-3) \cr
$X(^{58}$Ni) &   2.21(-5) &   2.56(-3) &   8.26(-3) &   3.29(-2) &   5.77(-2) &   8.26(-2) &   1.58(-1) &   2.34(-1) &   4.94(-1) &   4.31(-1) \cr
$P_{44}$     &      0.238 &      1.037 &      1.223 &      1.258 &      1.300 &      1.349 &      1.543 &      1.818 &      6.120 &     41.053 \cr
$P_{57}$     &      0.123 &      0.878 &      1.994 &      2.210 &      2.454 &      2.675 &      3.429 &      4.266 &     10.787 &     22.010 \cr
$P_{58}$     &      0.003 &      0.243 &      0.315 &      1.304 &      2.408 &      3.614 &      8.171 &     14.825 &    119.333 &   3732.058 \cr
\hline
\multicolumn{11}{c}{Point 4: Rotating Two-dimensional Explosion Model,  $T_{9p}= 6.5,  \rho_{p}=1.0\times 10^7$, S$_{\rm rad} = 9.14,  \tau_{\rm HD} = 0.14$ s} \\
$X(\alpha$)  &   1.03(-1) &   1.06(-1) &   1.05(-1) &   1.04(-1) &   1.03(-1) &   1.02(-1) &   9.77(-2) &   9.39(-2) &   8.14(-2) &   6.95(-2) \cr
$X(^{40}$Ca) &   2.00(-5) &   4.54(-4) &   4.32(-4) &   4.03(-4) &   3.81(-4) &   3.62(-4) &   3.15(-4) &   2.75(-4) &   1.74(-4) &   8.54(-5) \cr
$X(^{44}$Ti) &   3.11(-5) &   1.34(-4) &   1.90(-4) &   1.83(-4) &   1.77(-4) &   1.70(-4) &   1.51(-4) &   1.34(-4) &   8.76(-5) &   4.87(-5) \cr
$X(^{56}$Ni) &   8.50(-1) &   8.40(-1) &   8.21(-1) &   7.94(-1) &   7.67(-1) &   7.39(-1) &   6.56(-1) &   5.74(-1) &   3.04(-1) &   4.68(-2) \cr
$X(^{57}$Ni) &   4.15(-3) &   1.13(-2) &   2.79(-2) &   3.07(-2) &   3.32(-2) &   3.55(-2) &   4.07(-2) &   4.40(-2) &   4.30(-2) &   1.98(-2) \cr
$X(^{58}$Ni) &   7.02(-5) &   9.96(-4) &   1.14(-2) &   3.58(-2) &   6.04(-2) &   8.51(-2) &   1.60(-1) &   2.37(-1) &   5.03(-1) &   7.79(-1) \cr
$P_{44}$     &      0.029 &      0.119 &      0.173 &      0.173 &      0.172 &      0.172 &      0.172 &      0.175 &      0.216 &      0.777 \cr
$P_{57}$     &      0.267 &      0.571 &      1.445 &      1.648 &      1.839 &      2.044 &      2.649 &      3.268 &      6.017 &     18.011 \cr
$P_{58}$     &      0.021 &      0.183 &      0.317 &      1.029 &      1.790 &      2.624 &      5.566 &      9.408 &     37.801 &    379.032 \cr
\hline
\multicolumn{11}{l}{ $^{1}$ All results assume our ''recommended'' rates for $^{40}$Ca($\alpha,\gamma$)$^{44}$Ti and $^{44}$Ti($\alpha,p$)$^{47}$V.}\\
\label{nuc_survey}
\end{longtable}
\end{center}

\begin{center}
\begin{longtable}{ccccccccccc}
\multicolumn{11}{c}{Table 6. Nucleosynthesis Survey$^1$ ({\it continued})} \\
\hline \hline
$Y_e$        &     0.5050 &     0.5000 &     0.4995 &     0.4990 &     0.4985 &     0.4980 &     0.4965 &     0.4950 &     0.4900 &     0.4850  \\
$\eta$       &    -0.0100 &     0.0000 &     0.0010 &     0.0020 &     0.0030 &     0.0040 &     0.0070 &     0.0100 &     0.0200 &     0.0300  \\
\hline
\multicolumn{11}{c}{Point 5: Rotating Two-dimensional Explosion Model,  $T_{9p}= 5.5,  \rho_{p}=1.0\times 10^7$,   S$_{\rm rad} = 5.54,    \tau_{\rm HD} = 0.14$ s} \\
$X(\alpha$)  &   6.85(-2) &   7.23(-2) &   7.17(-2) &   7.07(-2) &   6.96(-2) &   6.85(-2) &   6.53(-2) &   6.21(-2) &   5.23(-2) &   4.36(-2) \cr
$X(^{40}$Ca) &   1.13(-5) &   3.68(-4) &   3.56(-4) &   3.29(-4) &   3.08(-4) &   2.90(-4) &   2.46(-4) &   2.10(-4) &   1.23(-4) &   6.11(-5) \cr
$X(^{44}$Ti) &   1.61(-5) &   8.16(-5) &   1.62(-4) &   1.54(-4) &   1.47(-4) &   1.41(-4) &   1.22(-4) &   1.05(-4) &   6.34(-5) &   3.43(-5) \cr
$X(^{56}$Ni) &   8.77(-1) &   8.74(-1) &   8.54(-1) &   8.27(-1) &   7.99(-1) &   7.71(-1) &   6.87(-1) &   6.04(-1) &   3.31(-1) &   6.85(-2) \cr
$X(^{57}$Ni) &   7.18(-3) &   9.75(-3) &   2.59(-2) &   2.92(-2) &   3.22(-2) &   3.48(-2) &   4.08(-2) &   4.45(-2) &   4.47(-2) &   2.40(-2) \cr
$X(^{58}$Ni) &   1.86(-4) &   8.99(-4) &   1.18(-2) &   3.54(-2) &   5.92(-2) &   8.33(-2) &   1.57(-1) &   2.32(-1) &   4.95(-1) &   7.73(-1) \cr
$P_{44}$     &      0.015 &      0.070 &      0.142 &      0.139 &      0.137 &      0.136 &      0.132 &      0.130 &      0.143 &      0.373 \cr
$P_{57}$     &      0.451 &      0.474 &      1.291 &      1.502 &      1.717 &      1.928 &      2.527 &      3.125 &      5.741 &     14.926 \cr
$P_{58}$     &      0.056 &      0.157 &      0.316 &      0.977 &      1.685 &      2.467 &      5.201 &      8.771 &     34.144 &    257.353 \cr
\hline
\multicolumn{11}{c}{Point 6: Rotating Two-dimensional Explosion Model,  $T_{9p}= 4.7,  \rho_{p}=1.0\times 10^7$,   S$_{\rm rad} = 3.45,    \tau_{\rm HD} = 0.14$ s} \\
$X(\alpha$)  &   5.67(-2) &   6.07(-2) &   6.04(-2) &   5.97(-2) &   5.90(-2) &   5.84(-2) &   5.65(-2) &   5.48(-2) &   4.99(-2) &   4.58(-2) \cr
$X(^{40}$Ca) &   1.36(-5) &   4.91(-4) &   5.04(-4) &   4.67(-4) &   4.42(-4) &   4.20(-4) &   3.71(-4) &   3.32(-4) &   2.36(-4) &   1.46(-4) \cr
$X(^{44}$Ti) &   1.19(-5) &   6.90(-5) &   2.42(-4) &   2.33(-4) &   2.24(-4) &   2.17(-4) &   1.96(-4) &   1.79(-4) &   1.34(-4) &   9.15(-5) \cr
$X(^{56}$Ni) &   8.68(-1) &   8.76(-1) &   8.56(-1) &   8.28(-1) &   7.99(-1) &   7.71(-1) &   6.86(-1) &   6.02(-1) &   3.26(-1) &   6.19(-2) \cr
$X(^{57}$Ni) &   1.14(-2) &   8.83(-3) &   2.47(-2) &   2.85(-2) &   3.18(-2) &   3.46(-2) &   4.09(-2) &   4.46(-2) &   4.42(-2) &   2.28(-2) \cr
$X(^{58}$Ni) &   4.46(-4) &   8.06(-4) &   1.05(-2) &   3.22(-2) &   5.43(-2) &   7.66(-2) &   1.45(-1) &   2.15(-1) &   4.59(-1) &   7.14(-1) \cr
$P_{44}$     &      0.011 &      0.059 &      0.210 &      0.210 &      0.209 &      0.209 &      0.213 &      0.222 &      0.306 &      1.102 \cr
$P_{57}$     &      0.720 &      0.428 &      1.229 &      1.464 &      1.685 &      1.912 &      2.532 &      3.159 &      5.753 &     15.650 \cr
$P_{58}$     &      0.139 &      0.143 &      0.281 &      0.888 &      1.550 &      2.271 &      4.826 &      8.159 &     32.124 &    262.195 \cr
\hline
\multicolumn{11}{c}{Point 7: Rotating Two-dimensional MHD Model (CHASM region),  $T_{9p}= 6.5,  \rho_{p}=1.0\times 10^8$,   S$_{\rm rad} = 0.91,  \tau_{\rm HD} = 0.04$ s} \\
$X(\alpha$)  &   1.01(-2) &   1.31(-2) &   1.20(-2) &   1.08(-2) &   9.65(-3) &   8.51(-3) &   5.25(-3) &   2.26(-3) &   1.81(-6) &   8.95(-7) \cr
$X(^{40}$Ca) &   7.43(-8) &   3.61(-5) &   2.64(-5) &   1.94(-5) &   1.41(-5) &   9.95(-6) &   2.68(-6) &   2.88(-7) &   4.06(-8) &   2.35(-7) \cr
$X(^{44}$Ti) &   1.22(-7) &   1.15(-5) &   1.11(-5) &   8.25(-6) &   5.97(-6) &   4.19(-6) &   1.08(-6) &   1.06(-7) &   1.98(-9) &   6.31(-9) \cr
$X(^{56}$Ni) &   8.67(-1) &   9.40(-1) &   9.14(-1) &   8.87(-1) &   8.59(-1) &   8.32(-1) &   7.52(-1) &   6.71(-1) &   3.93(-1) &   1.28(-1) \cr
$X(^{57}$Ni) &   1.06(-2) &   1.34(-2) &   2.30(-2) &   2.84(-2) &   3.30(-2) &   3.69(-2) &   4.57(-2) &   5.15(-2) &   4.88(-2) &   3.07(-2) \cr
$X(^{58}$Ni) &   7.53(-4) &   1.50(-4) &   1.92(-2) &   4.21(-2) &   6.61(-2) &   9.11(-2) &   1.72(-1) &   2.60(-1) &   4.30(-1) &   5.61(-1) \cr
$P_{44}$     &      0.000 &      0.009 &      0.009 &      0.007 &      0.005 &      0.004 &      0.001 &      0.000 &      0.000 &      0.000 \cr
$P_{57}$     &      1.327 &      0.607 &      1.069 &      1.365 &      1.642 &      1.891 &      2.596 &      3.265 &      5.272 &     10.388 \cr
$P_{58}$     &      1.244 &      0.064 &      0.478 &      1.082 &      1.760 &      2.496 &      5.209 &      8.837 &     24.888 &     99.029 \cr
\hline
\multicolumn{11}{c}{Point 8: Rotating Two-dimensional MHD Model,  $T_{9p}= 5.5,  \rho_{p}=1.0\times 10^8$,   S$_{\rm rad} = 0.55,    \tau_{\rm HD} = 0.04$ s} \\
$X(\alpha$)  &   9.27(-3) &   1.13(-2) &   1.07(-2) &   9.89(-3) &   9.16(-3) &   8.47(-3) &   6.57(-3) &   4.86(-3) &   4.69(-4) &   2.45(-6) \cr
$X(^{40}$Ca) &   1.39(-7) &   5.28(-5) &   4.50(-5) &   3.56(-5) &   2.84(-5) &   2.27(-5) &   1.12(-5) &   4.92(-6) &   1.59(-8) &   4.70(-10)\cr
$X(^{44}$Ti) &   2.14(-7) &   6.95(-6) &   2.06(-5) &   1.65(-5) &   1.33(-5) &   1.07(-5) &   5.20(-6) &   2.24(-6) &   5.93(-9) &   4.65(-11)\cr
$X(^{56}$Ni) &   7.86(-1) &   9.25(-1) &   9.01(-1) &   8.74(-1) &   8.47(-1) &   8.20(-1) &   7.39(-1) &   6.59(-1) &   3.89(-1) &   1.18(-1) \cr
$X(^{57}$Ni) &   1.07(-2) &   8.79(-3) &   2.10(-2) &   2.67(-2) &   3.14(-2) &   3.52(-2) &   4.39(-2) &   4.95(-2) &   5.37(-2) &   3.37(-2) \cr
$X(^{58}$Ni) &   1.40(-3) &   2.13(-4) &   1.47(-2) &   3.50(-2) &   5.64(-2) &   7.89(-2) &   1.52(-1) &   2.32(-1) &   5.45(-1) &   7.60(-1) \cr
$P_{44}$     &      0.000 &      0.006 &      0.017 &      0.014 &      0.012 &      0.010 &      0.005 &      0.003 &      0.000 &      0.000 \cr
$P_{57}$     &      1.505 &      0.404 &      0.992 &      1.298 &      1.575 &      1.828 &      2.521 &      3.193 &      5.890 &     12.301 \cr
$P_{58}$     &      2.660 &      0.086 &      0.373 &      0.914 &      1.516 &      2.197 &      4.685 &      8.031 &     31.974 &    146.341 \cr
\hline
\multicolumn{11}{l}{ $^{1}$ All results assume our ''recommended'' rates for $^{40}$Ca($\alpha,\gamma$)$^{44}$Ti and $^{44}$Ti($\alpha,p$)$^{47}$V.}\\ 
\end{longtable}
\end{center}

\begin{figure}
\begin{center}
\includegraphics[scale=.60]{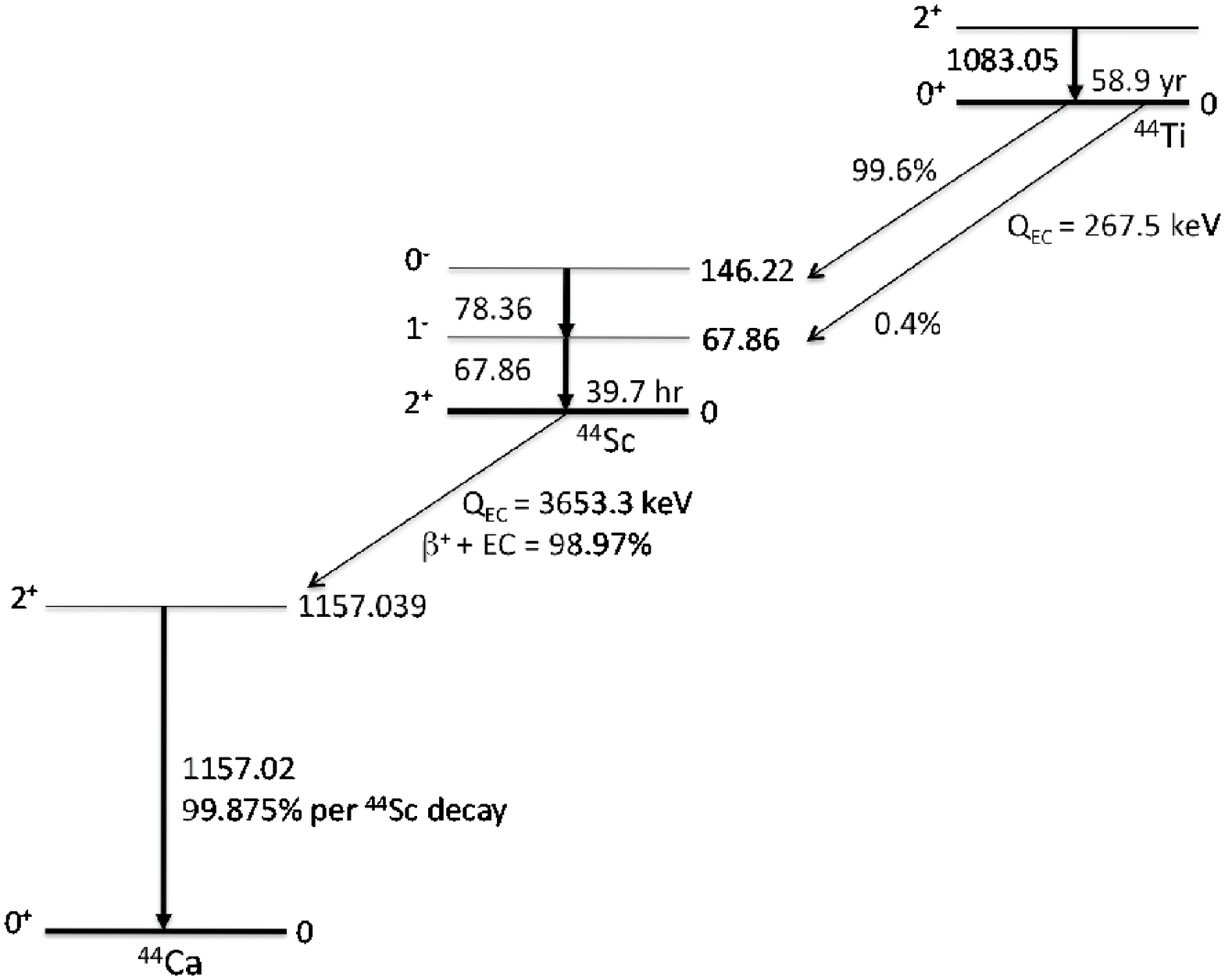}
\caption{\label{partialdecay}Partial decay schemes of $^{44}$Ti and its daughter $^{44}$Sc.}
\end{center}
\end{figure}

\clearpage
\begin{figure}
\begin{center}
\includegraphics[scale=.30]{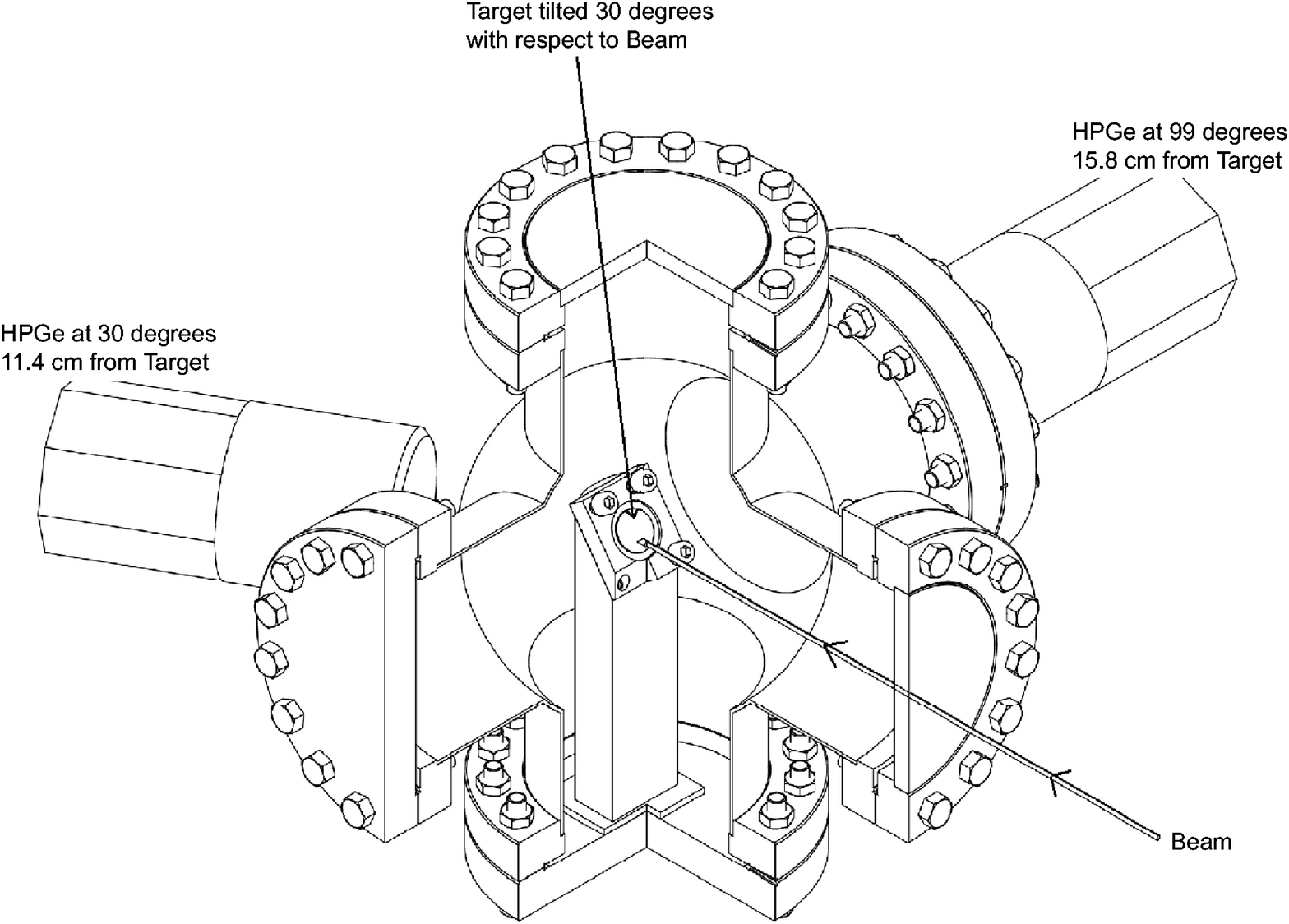}
\caption{\label{chamber} A Schematic of the target chamber used in the experiment.} 
\end{center}
\end{figure}

\clearpage
\begin{figure}
\begin{center}
\includegraphics[scale=.80,angle=90]{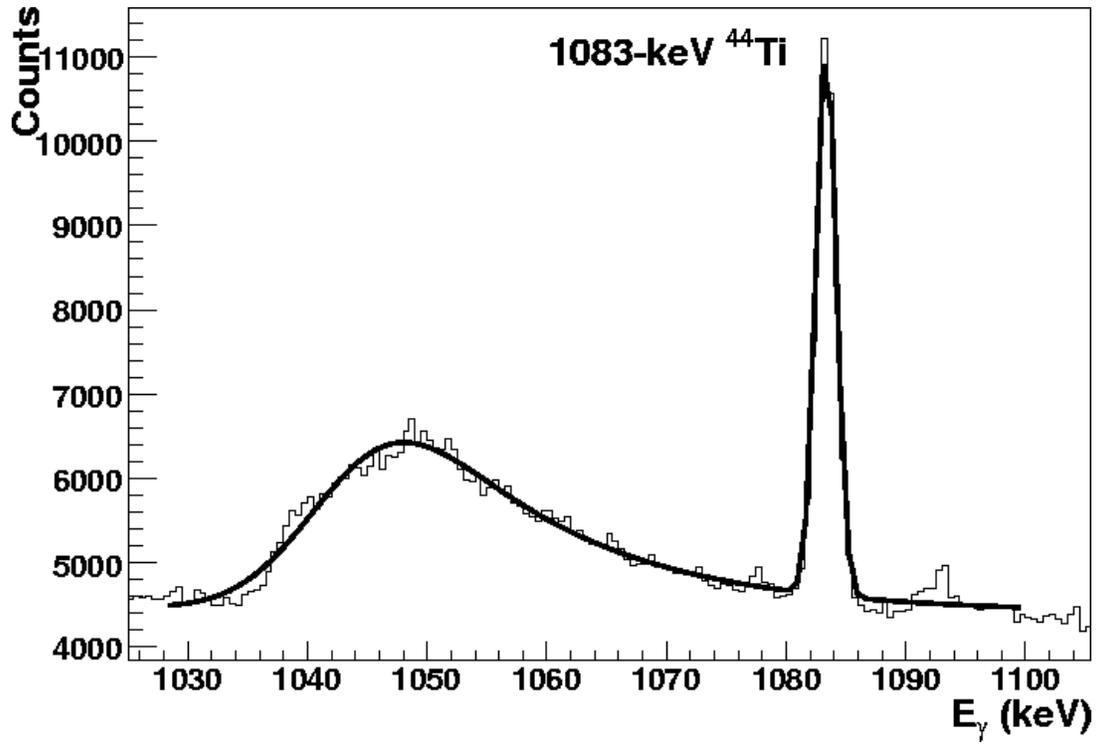}
\caption{Partial HPGe $\gamma$-ray spectra at $E_{\alpha}$ = 5.36 for the detector at 
99$^\circ$ with a simultaneous fit to the 1039 keV $^{70}$Ge and 1083 keV $^{44}$Ti $\gamma$-rays. }
\label{partialspectra}
\end{center}
\end{figure}

\clearpage
\begin{figure}
\begin{center}
\includegraphics[scale=0.70]{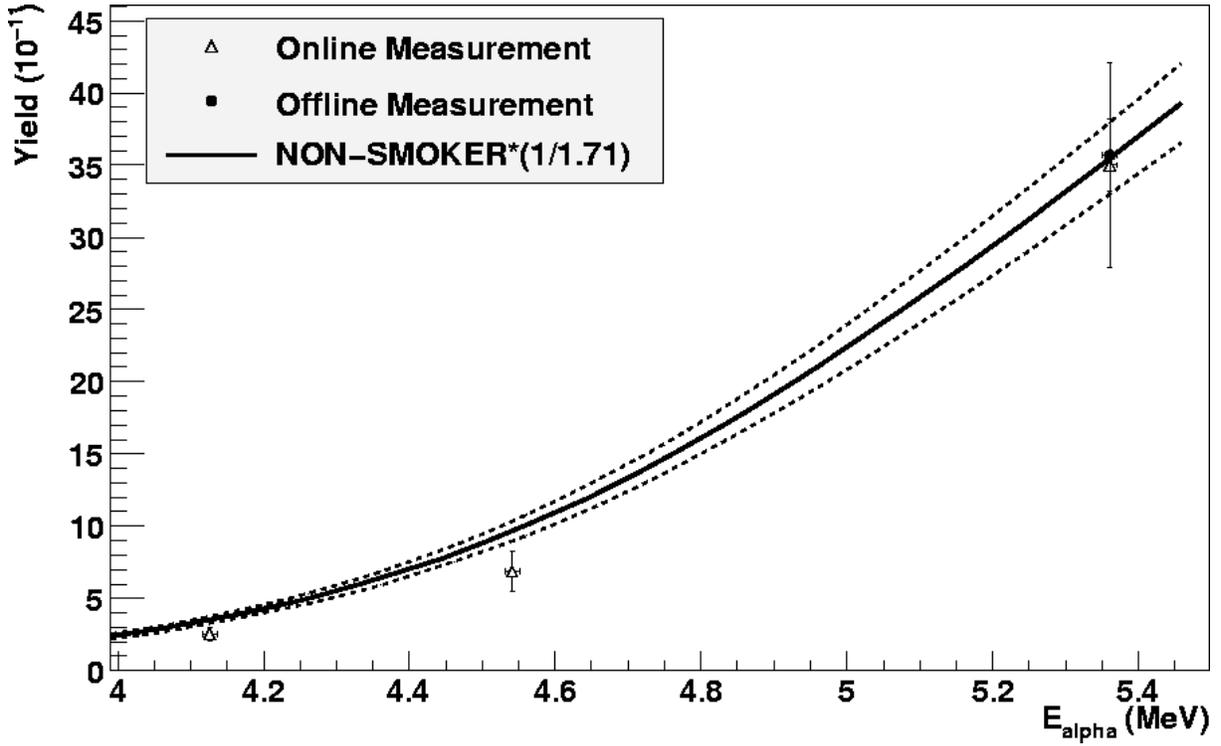}
\caption{Measured thick target yields compared to the calculated NON-SMOKER thick target yield. 
The online measurement is the measured yield for the 1083 keV $\gamma$-ray corrected upward by 20$\%$ (see the text). 
The $E_{\alpha}$ = 5.36 MeV offline counting data point and its error is plotted as measured. The NON-SMOKER cross section 
required a downward scaling of 1.71 to match the off-line data point. 
The dashed lines indicate the normalization needed to match the offline error bars.}
\label{thickyield}
\end{center}
\end{figure}

\clearpage
\begin{figure}
\begin{center}
\includegraphics[scale=0.80,angle=90]{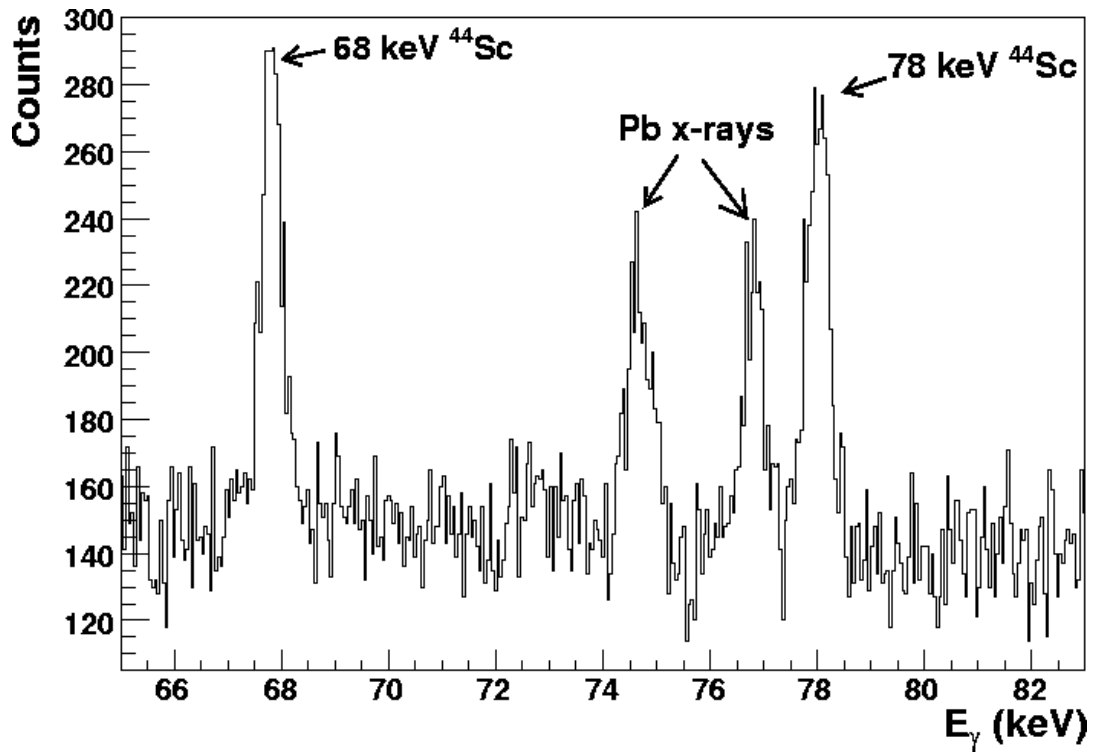}
\caption{$\gamma$-ray spectra observed in a two week low background count of the activated target bombarded at $E_\alpha$ = 5.36 MeV.}
\label{lepsspectra}
\end{center}
\end{figure}

\clearpage
\begin{figure}
\begin{center}
\includegraphics[scale=0.6]{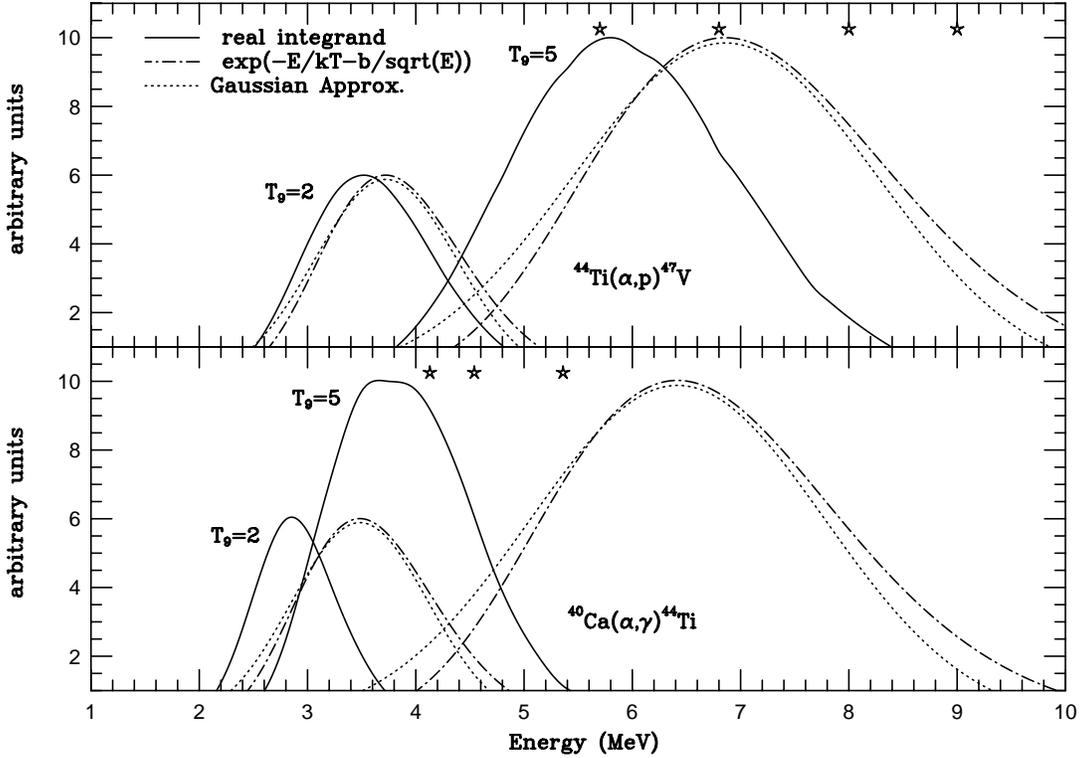}
\caption{The Effective Gamow windows for $^{40}$Ca($\alpha$,$\gamma$)$^{44}$Ti and $^{44}$Ti($\alpha,p$)$^{47}$V at $T_9=2$ and $T_9=5$. 
Shown are the true Gamow window as given
by the real integrand of the reaction rate formula (solid line type, see Eq.\ \ref{eq:reacrate}), the product of the high-energy tail
of the Maxwell-Boltzmann distribution with the simple Coulomb barrier penetration factor (denoted as ``exp(-E/kT-b/sqrt(E)), dot-dash line type)'' 
and the usual Gaussian approximation to the latter (``Gauss approx'', dotted line type). Stars denote the energy values at which experimental
data were measured in our effort and from \cite{sonz00}.
The Gaussian approximation is widely used to estimate the Gamow window but predicts it at too high an energy, especially for the capture reaction. 
For purposes of illustration, all curves were renormalized to the arbitrary unit scale shown, for $T_9=5$ all
three quantities peak at 10,
for $T_9=2$ they peak at 6.}
\label{fig:gamowca40ag}
\end{center}
\end{figure}

\clearpage
\begin{figure}
\begin{center}
\includegraphics[scale=0.6]{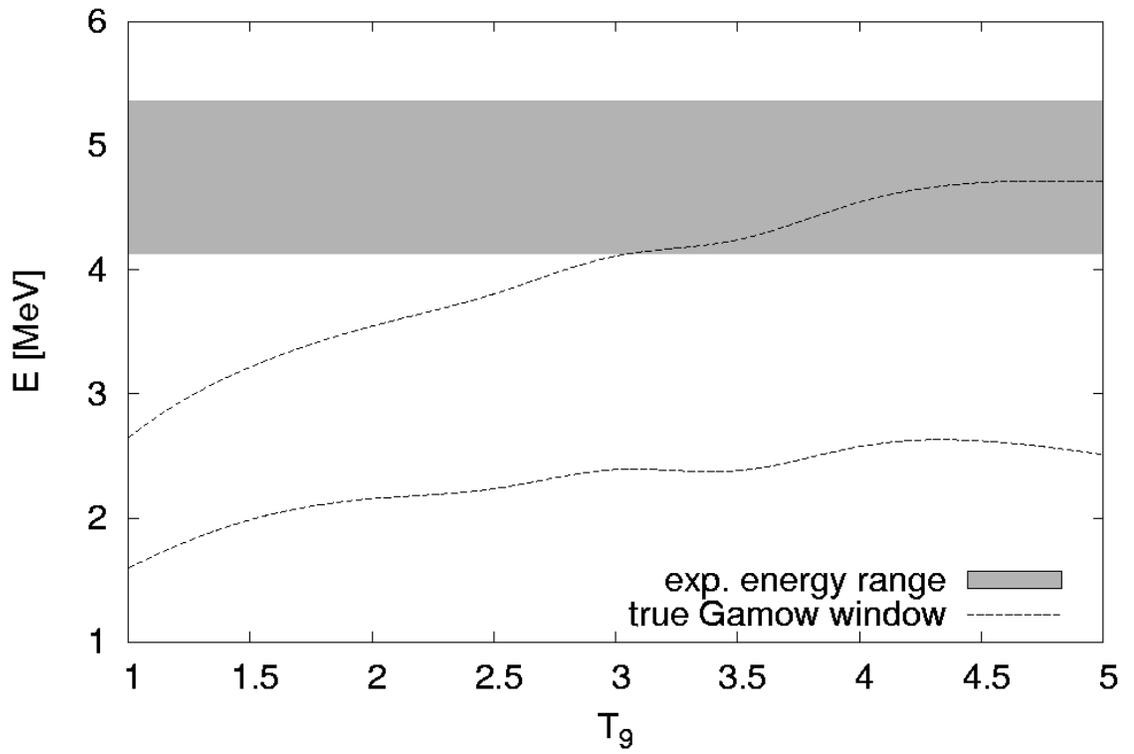}
\caption{Comparison of the experimental energy range for $^{40}$Ca($\alpha$,$\gamma$)$^{44}$Ti with the true Gamow window given
by the actual maximal contributions to the reaction rate integral. The experimental data do not contribute significantly to the rate
in the temperature range relevant to $^{44}$Ti synthesis ($2.0 \leq {\rm T}_9 \leq 4.0$).}
\label{fig:gamow2exp}
\end{center}
\end{figure}

\clearpage
\begin{figure}
\begin{center}
\includegraphics[scale=0.6]{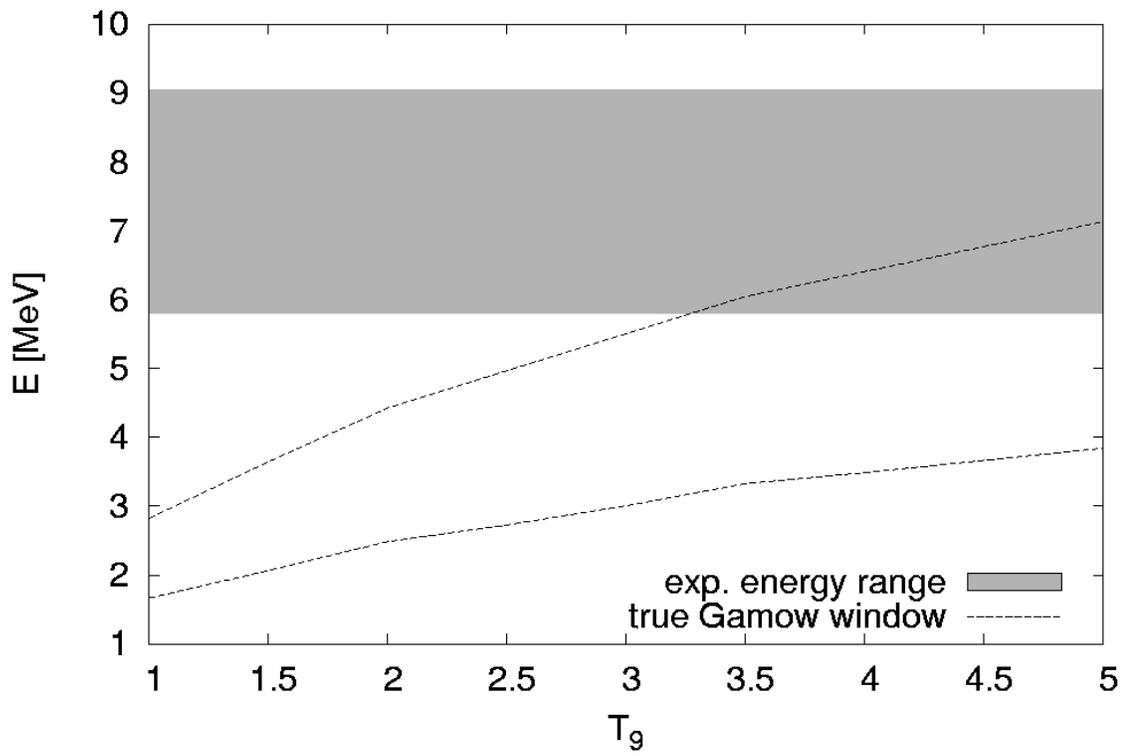}
\caption{Same as Figure \ref{fig:gamow2exp} but for the $^{44}$Ti($\alpha$,$p$)$^{47}$V reaction.}
\label{fig:gamow2expb}
\end{center}
\end{figure}

\clearpage
\begin{figure}
\begin{center}
\includegraphics[scale=0.6,angle=0]{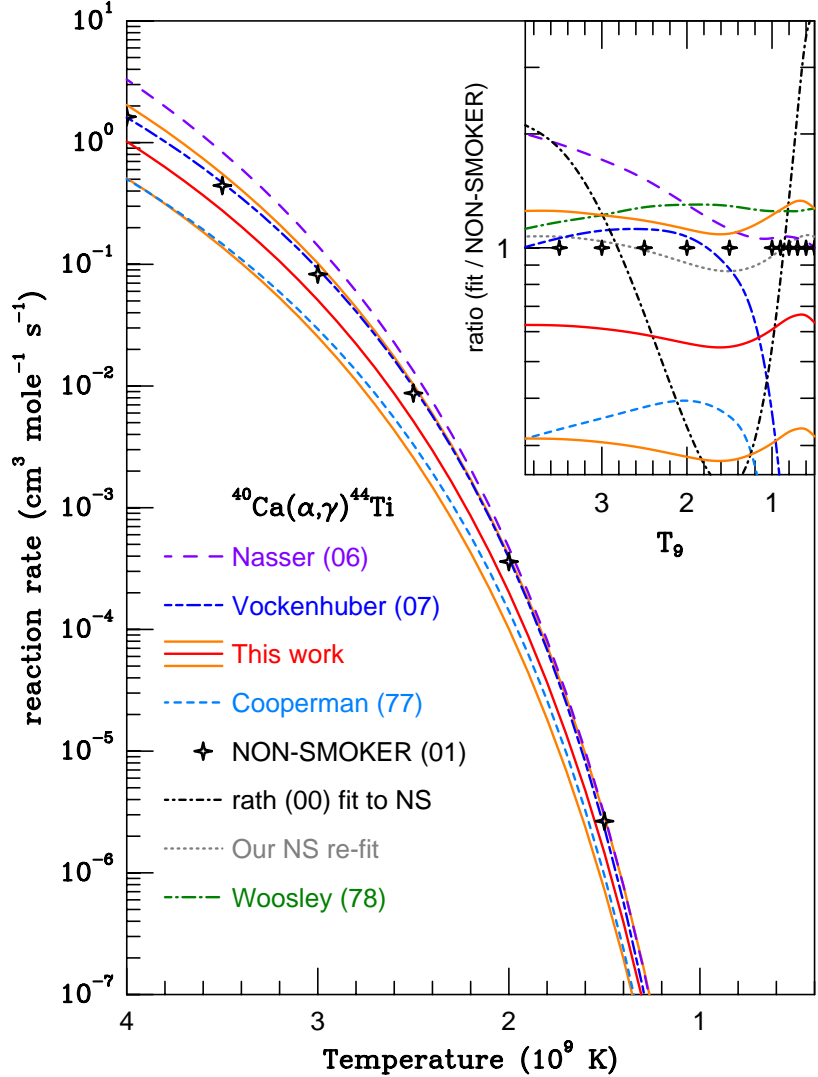}
\caption{
Fits to reaction rates for $^{40}$Ca($\alpha$,$\gamma$)$^{44}$Ti considered in this study.
Solid and dashed line types denote experimental rates from various authors, 
dotted and dot-dash line types are from various theory efforts.
The tabulated NON-SMOKER theory rate \citep{rauscher01} is denoted by crosses.
Our recommended rate (and its error, $\pm \times 2$) are indicated as solid (lighter) lines.
The inset illustrates the ratio of each reaction rate to the NON-SMOKER rate, including the
(dot-short-dash) original fit \citep{rauscher00b}, our attempt (dotted) to refit the tabulated NON-SMOKER rate,
and the theory rate (dot-long-dash) of \cite{wfhz78}. 
}
\label{ca40ag_rates}
\end{center}
\end{figure}

\clearpage
\begin{figure}
\begin{center}
\includegraphics[scale=0.6,angle=0]{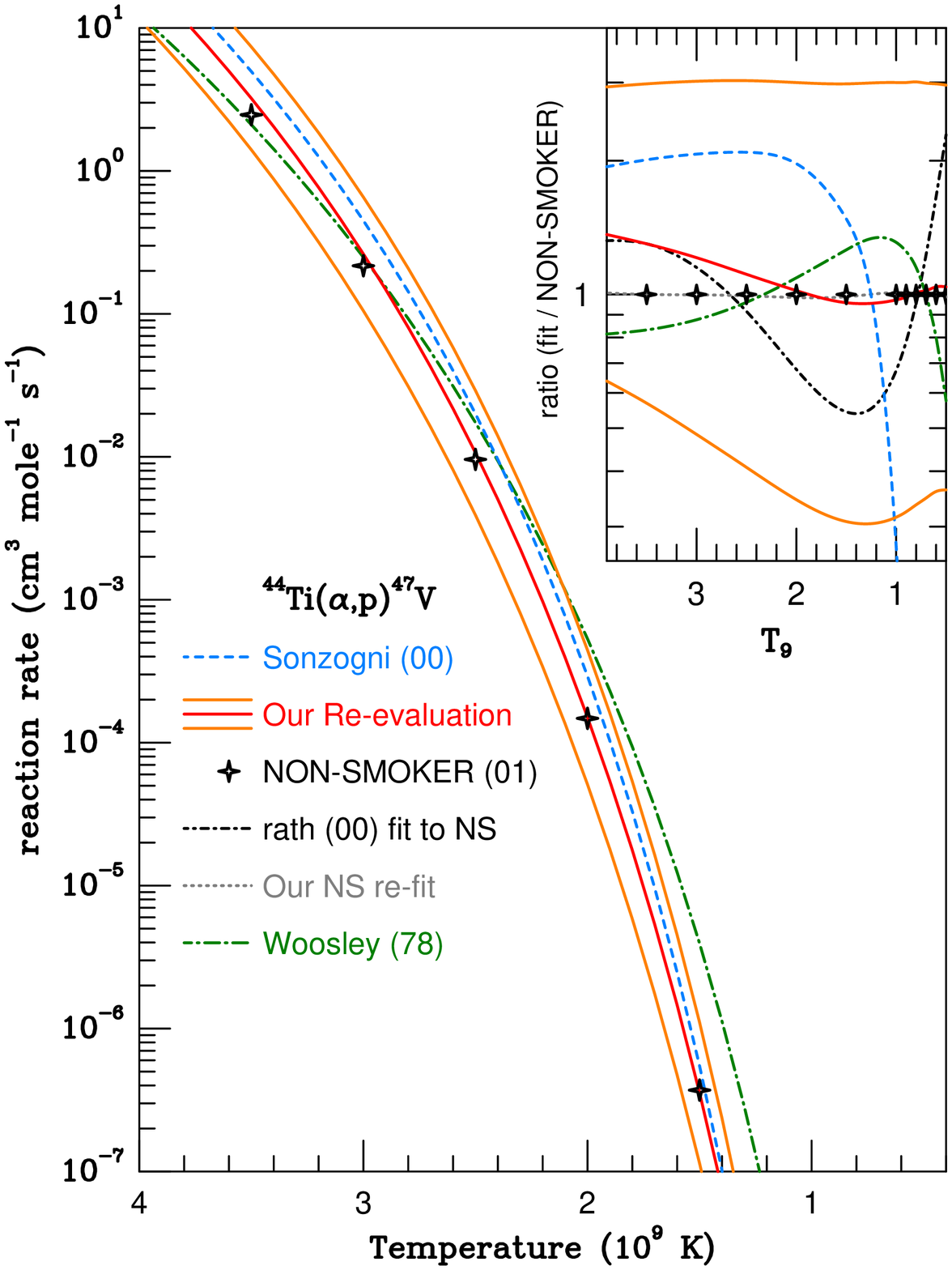}
\caption{
Fits to reaction rates for $^{44}$Ti($\alpha$,$p$)$^{47}$V considered in this study.
Solid line type denotes our re-evaluation and its likely error ($\pm \times 3$) 
of the lone experimental rate (dashed-line type) of \cite{sonz00}, 
while the tabulated NON-SMOKER theory rate \citep{rauscher01} is denoted by crosses.
The inset illustrates the ratio of each reaction rate to the NON-SMOKER rate, including the 
(dot-short-dash) original fit \citep{rauscher00b}, our attempt (dotted) to refit the tabulated NON-SMOKER rate, and
the theory rate (dot-long-dash) from \cite{wfhz78}.   
}
\label{ti44ap_rates}
\end{center}
\end{figure}

\clearpage
\begin{figure}
\begin{center}
\includegraphics[scale=0.7,angle=-90]{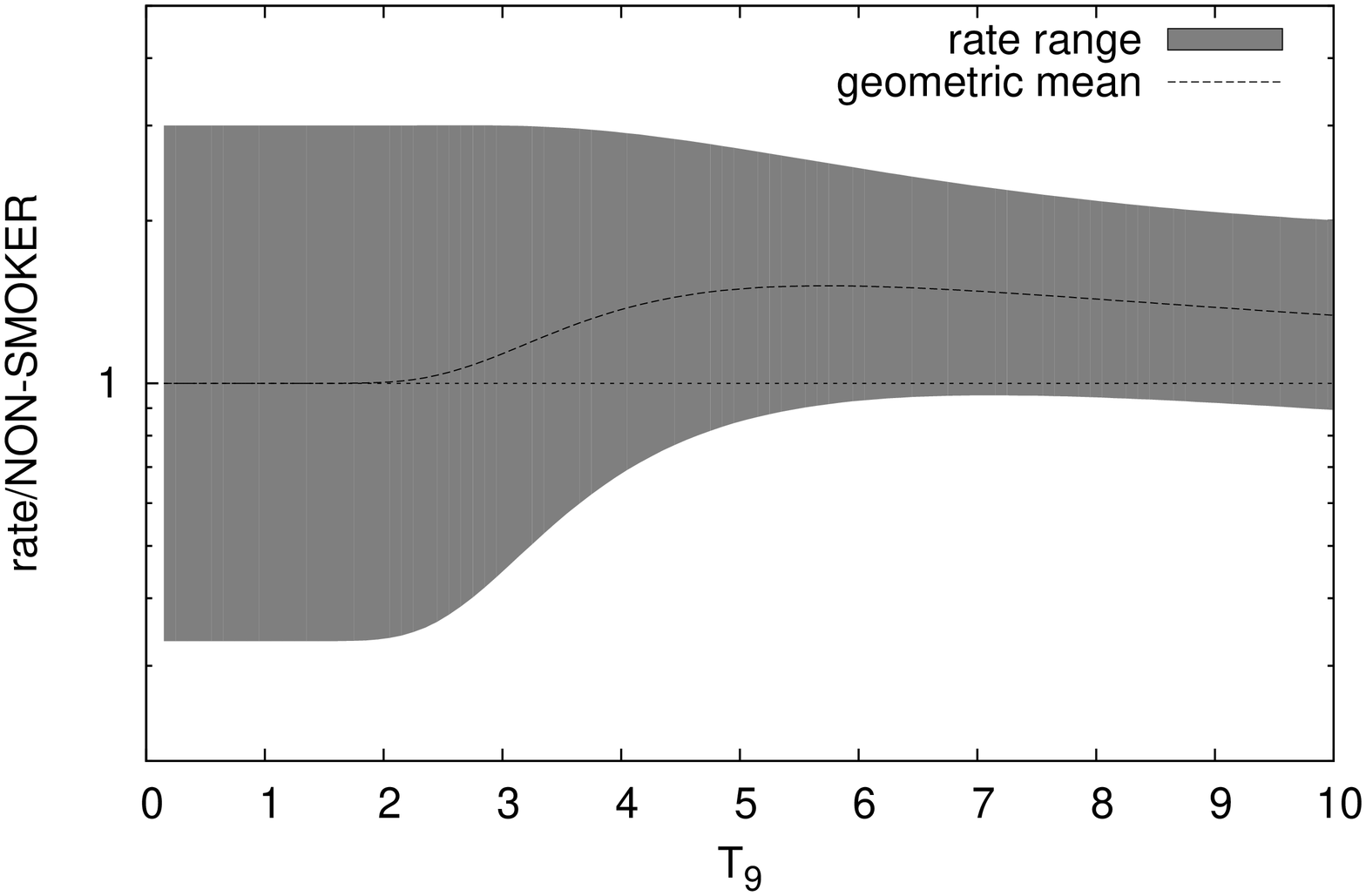}
\caption{Re-evaluated $^{44}$Ti($\alpha$,$p$)$^{47}$V rate, based on data by \cite{sonz00}: Ratio of the new rate to the
NON-SMOKER rate \citep{rauscher01}. The estimated range of the rate is derived from the theoretical and experimental errors
(see the text for details). The geometric mean of the upper and lower limit is used as a recommended rate in this work. The impact of the
\cite{sonz00} data manifests itself in the deviation of the geometric mean from unity. Within errors, the new rate is compatible
with the NON-SMOKER rate.}
\label{fig:newsonz}
\end{center}
\end{figure}

\clearpage
\begin{figure}
\begin{center}
\includegraphics[scale=0.7]{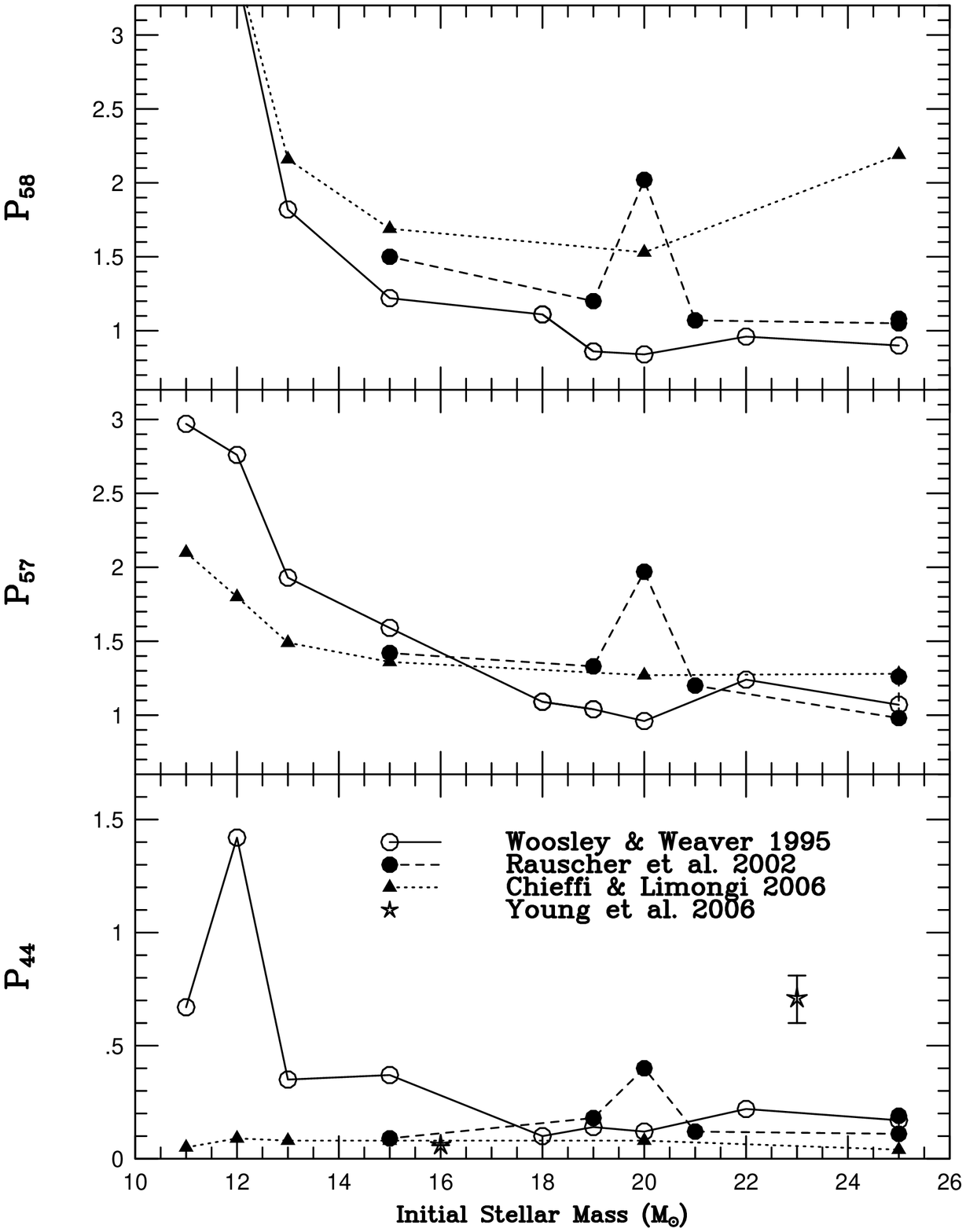}
\caption{
Normalized production factors for $^{44}$Ti, $^{57}$Ni and $^{58}$Ni versus initial stellar mass derived
from nucleosynthetic yields of SN models \citep{woosley95,rauscher02,cl06,young06}.
}
\label{sn2_npf_sol}
\end{center}
\end{figure}

\clearpage
\begin{figure}
\begin{center}
\includegraphics[scale=0.6]{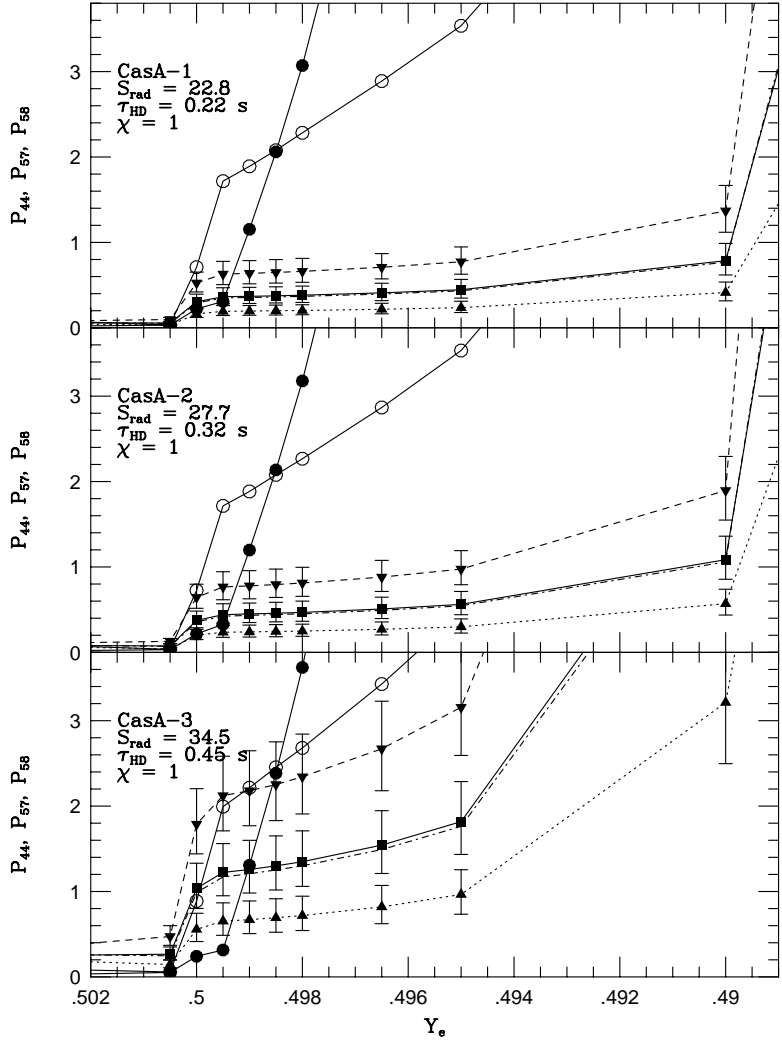}
\caption{
Normalized production factors for $^{44}$Ti, $^{57}$Ni, and $^{58}$Ni versus $Y_e$ for adiabatic
freeze outs from peak conditions CasA $1-3$ in Table \ref{conditions}.
All points were drawn from a model for Cassiopia A \citep{mag08}.
Each central point represents a calculation that utilizes our recommended "semi-experimental" reaction rate for
$^{40}$Ca($\alpha,\gamma$)$^{44}$Ti and three choices of $^{44}$Ti($\alpha,p$)$^{47}$V reaction rate
from our re-evaluation of the experimental effort of \cite{sonz00}: our "recommended" $^{44}$Ti($\alpha,p$)$^{47}$V rate 
(solid line, filled squares), and its upper and lower bound (denoted by filled triangles with dotted and dashed line styles, respectively).
Error bars on each point reflect the minimum and maximum deviations of $P_{44}$
due to variations of the six other $^{40}$Ca($\alpha,\gamma$)$^{44}$Ti reaction rates we considered.
The dot-short-dash line shows $^{44}$Ti synthesis using the original fits of \cite{rauscher00b}
for both $^{40}$Ca($\alpha,\gamma$)$^{44}$Ti and $^{44}$Ti($\alpha,p$)$^{47}$V.
For the two NPF's $P_{57}$ and $P_{58}$ (shown as open and filled circles, respectively)
an upper bound of 3 limits the allowed range of $Y_e$.
}
\label{CasA_npf}
\end{center}
\end{figure}

\clearpage
\begin{figure}
\begin{center}
\includegraphics[scale=0.7]{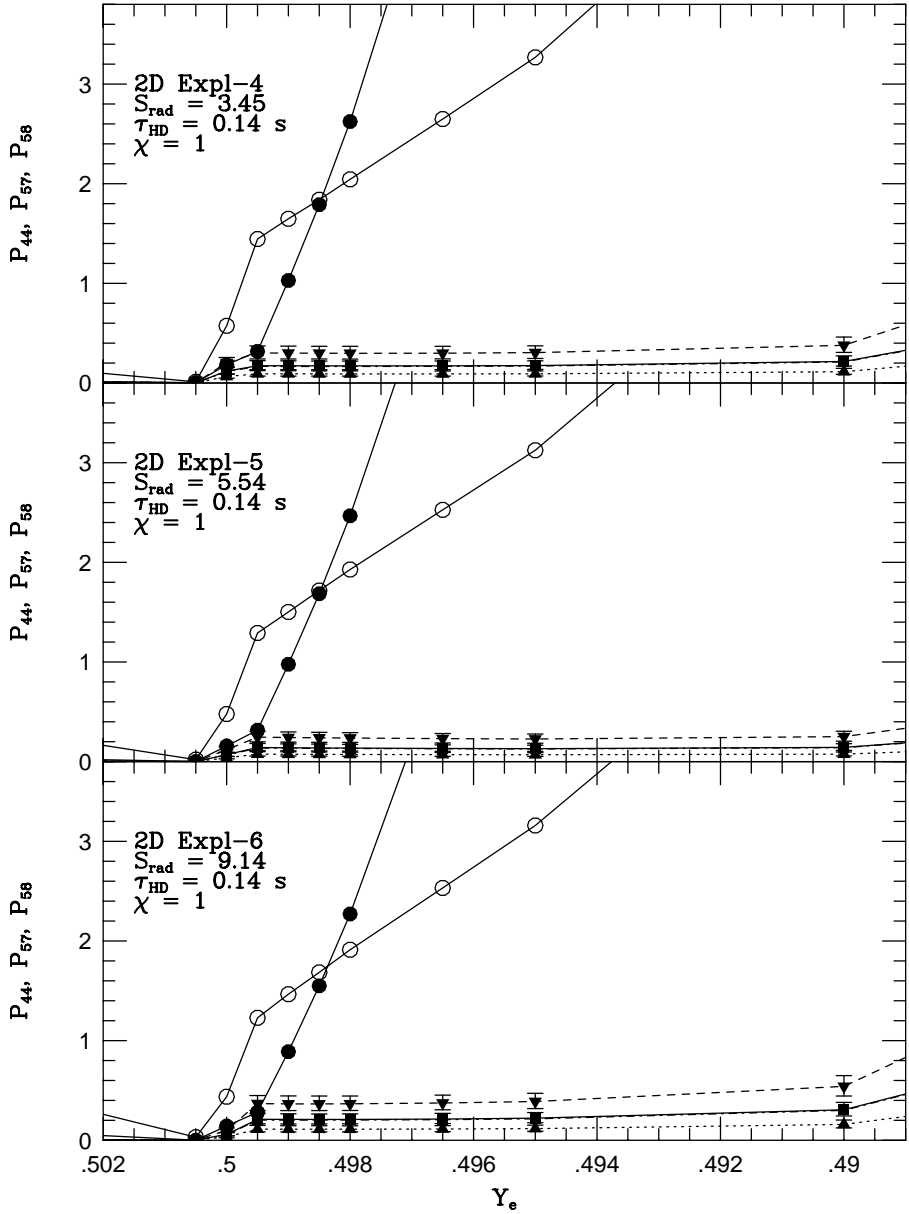}
\caption{
Same as Figure \ref{CasA_npf} but for peak conditions
$4-6$ in Table \ref{conditions} drawn from a rotating two-dimensional explosion model \citep{mag08}.
The production of $^{57,58}$Ni with respect to solar iron is nearly the same as in
the model for CasA, but the production of $^{44}$Ti is substantially lower in every case due to the lower
entropy experienced in these expansions.
}
\label{2DEx_npf}
\end{center}
\end{figure}

\clearpage
\begin{figure}
\begin{center}
\includegraphics[scale=0.7]{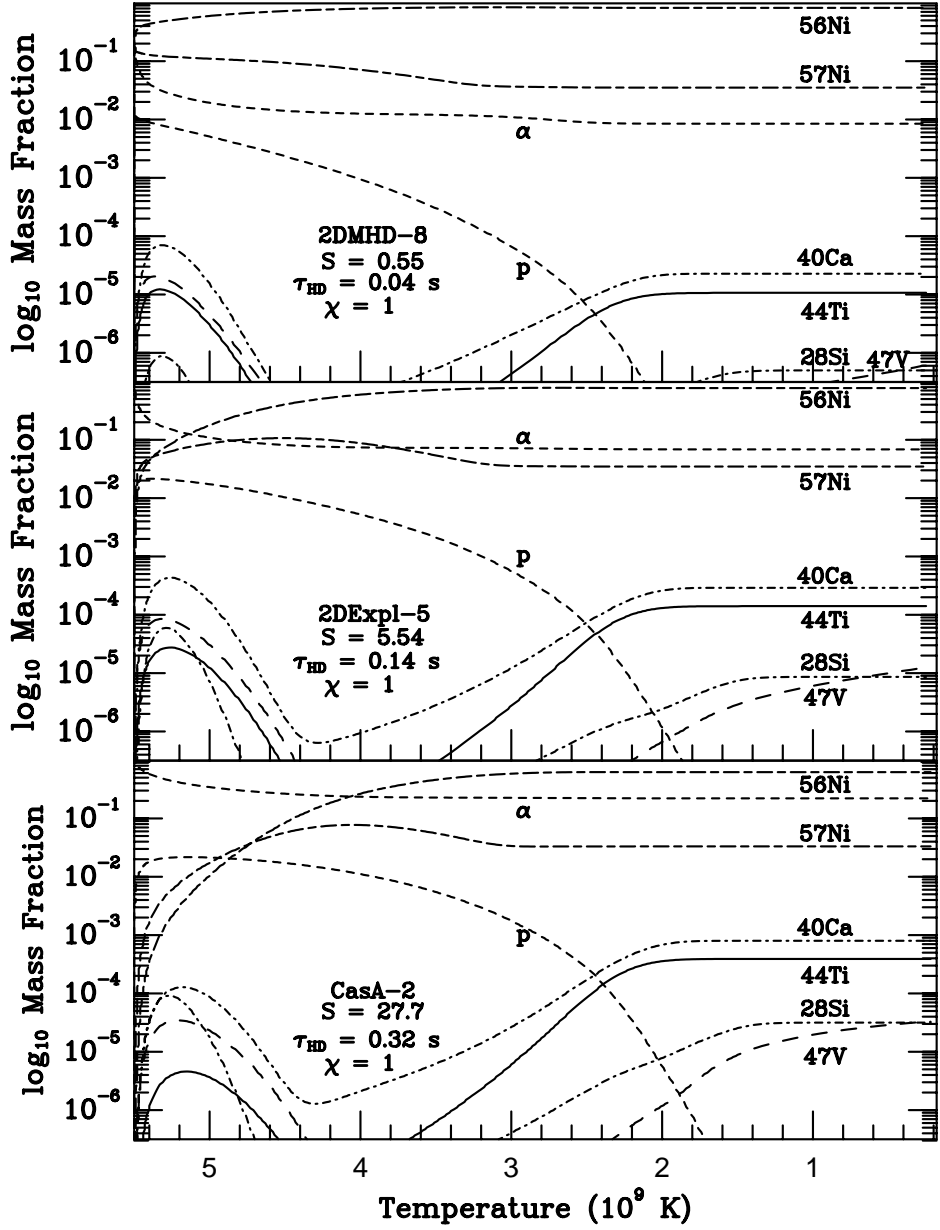}
\caption{
Evolution of select mass fractions versus temperature during freeze outs with identical peak temperatures ($T_{9p}=5.5$)
and electron fractions ($Y_e=0.4980$), but different intial peak densities, for three points in our survey.
The higher entropy expansions produce less $^{56}$Ni and have a higher $\alpha$-particle mass fraction at late times 
that facilitates the re-assembly of species below the iron group, including $^{40}$Ca and $^{44}$Ti.
}
\label{3panel_msfvst9}
\end{center}
\end{figure}

\clearpage
\begin{figure}
\begin{center}
\includegraphics[scale=0.6]{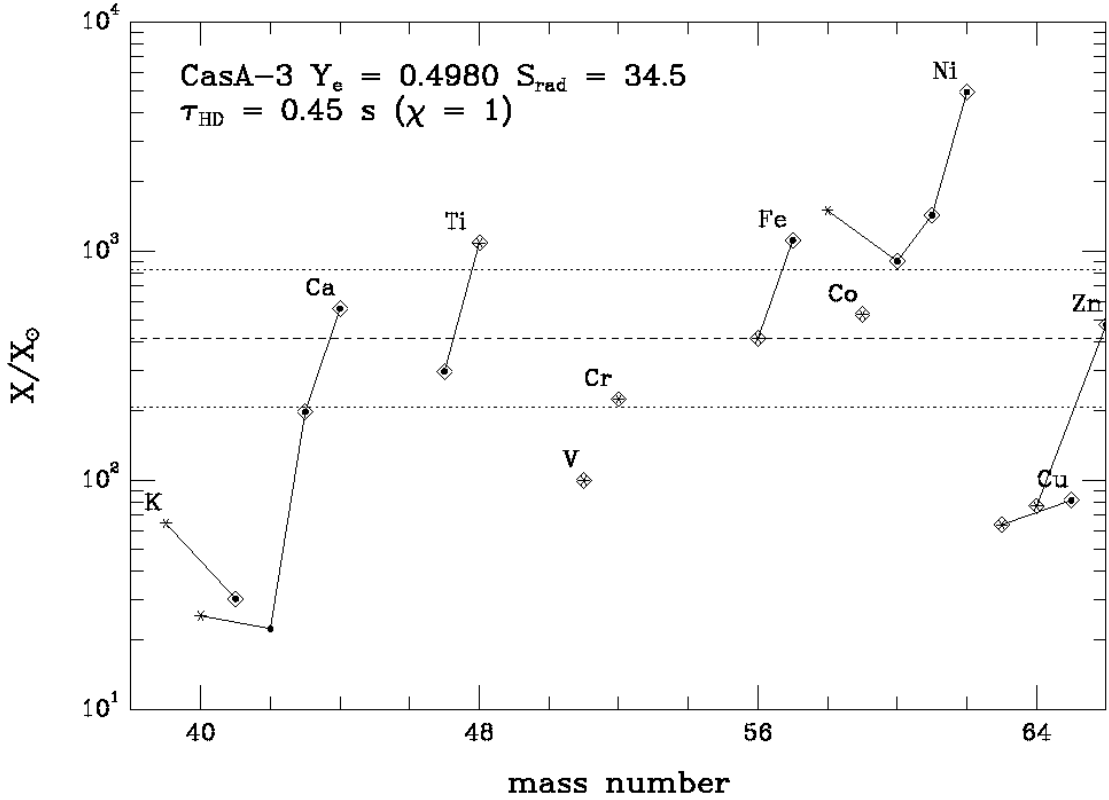}
\caption{
Production factors (final decayed mass fractions divided by their abundances in the Sun) representing the range of species produced 
in the $\alpha$-rich freeze-out from peak conditions for point CasA-3 in Table \ref{conditions} assuming
our ''recommended'' reaction 
rates for $^{40}$Ca($\alpha,\gamma$)$^{44}$Ti and $^{44}$Ti($\alpha,p$)$^{47}$V. This is
for the specific case of $Y_e=0.4980$ and $\chi = 1$. This is a fairly typical nucleosynthesis pattern 
over the range of allowed $Y_e$ for the $\alpha$-rich freeze outs in the CasA and two-dimensional explosion models we consider.
}
\label{3yep4980psi1ovr}
\end{center}
\end{figure}

\clearpage
\begin{figure}
\begin{center}
\includegraphics[scale=0.6]{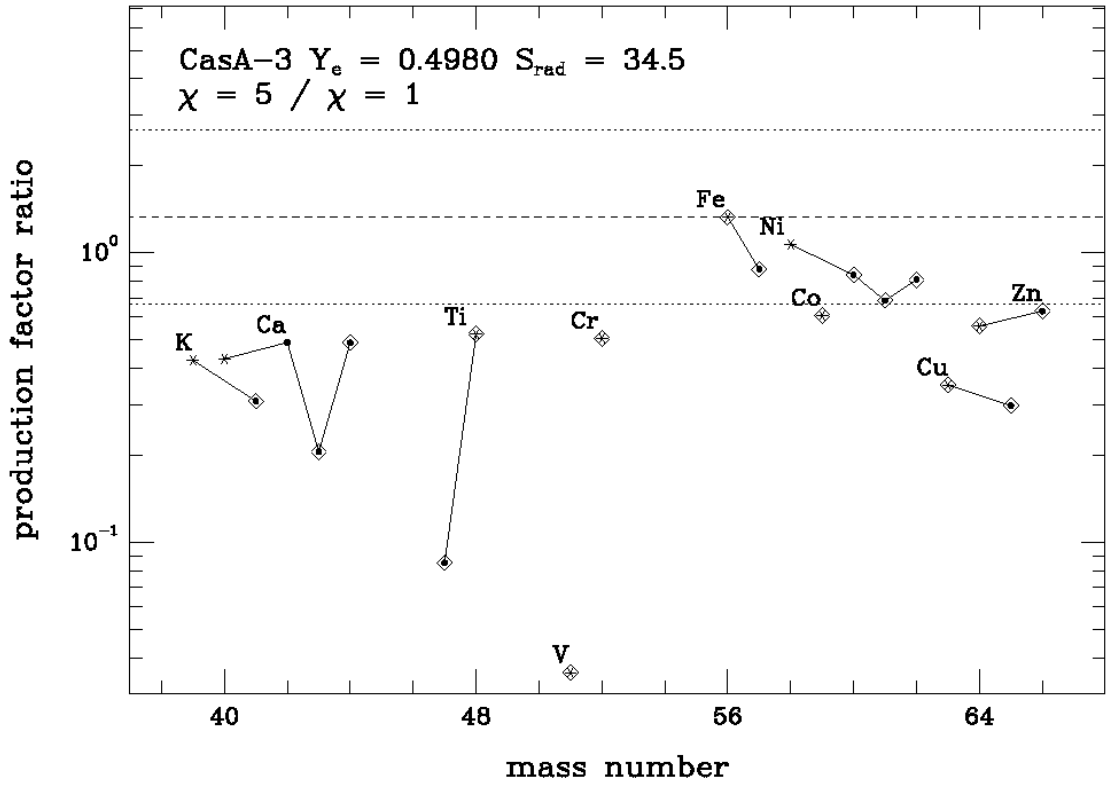}
\caption{Similar to Figure \ref{3yep4980psi1ovr},
this is the ratio of the production factors for 
two scalings on the hydrodynamic time-scale, default ($\chi=1$) and extended 
($\chi=5$). The species created in NSE are much less affected than
the lighter isotopes which are reassembled for $2\leq T_9 \leq 4$.
}
\label{ratioovr15}
\end{center}
\end{figure}

\clearpage
\begin{figure}
\begin{center}
\includegraphics[scale=0.6]{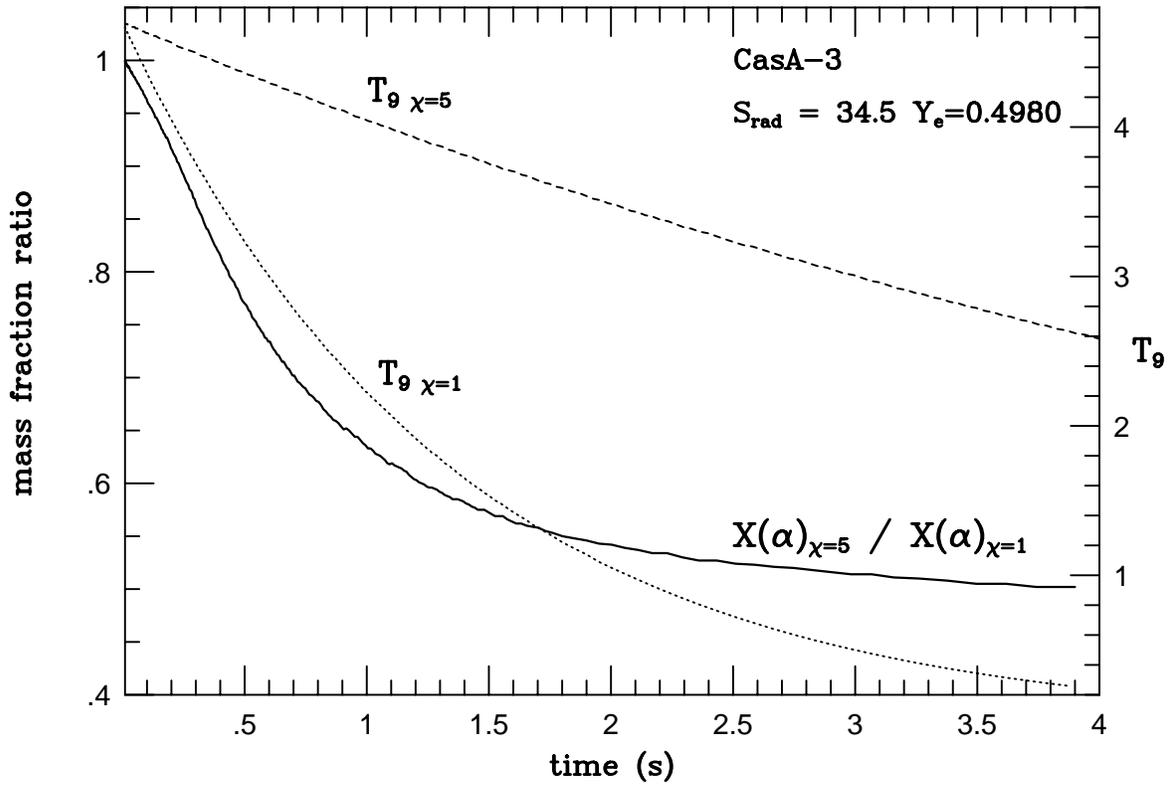}
\caption{Temperature evolution in the CasA-3 expansion assuming two
scalings on the hydrodynamic time-scale, default ($\chi=1$) and extended 
($\chi=5$). Also shown is the ratio of the $\alpha$-particle mass fraction.
}
\label{xalphapsi5db1}
\end{center}
\end{figure}

\clearpage
\begin{figure}
\begin{center}
\includegraphics[scale=0.6]{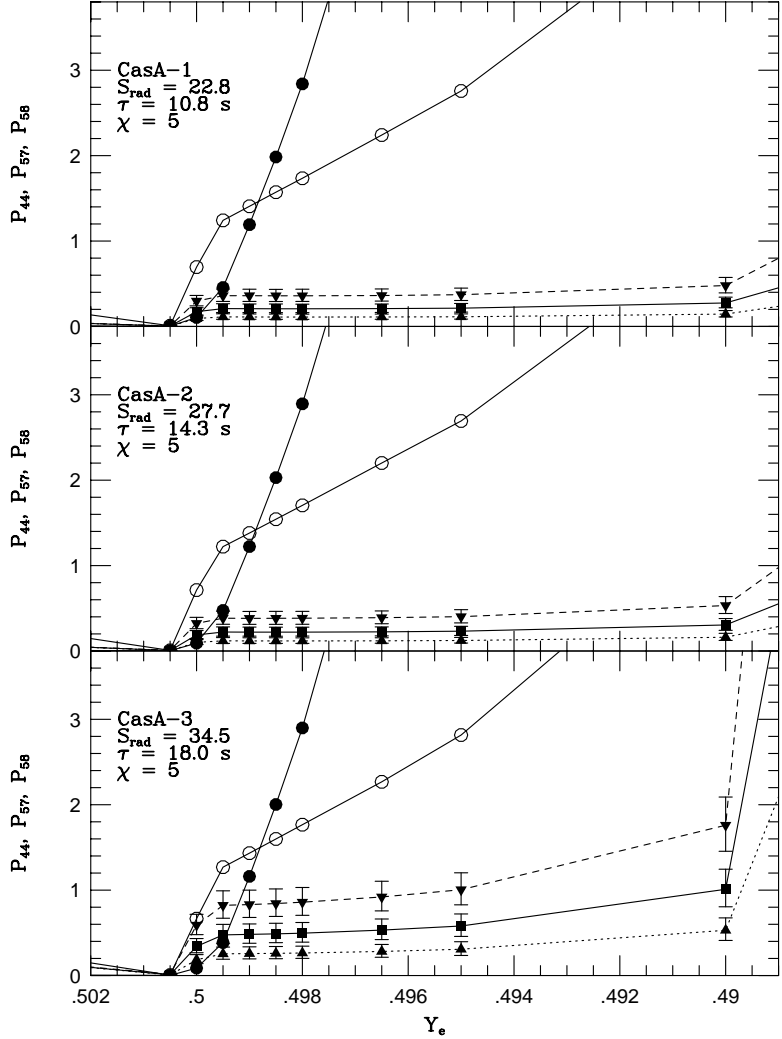}
\caption{
Normalized production factors for $^{44}$Ti, $^{57}$Ni, and $^{58}$Ni versus $Y_e$ for adiabatic
freeze outs from peak conditions CasA $1-3$ in Table \ref{conditions} but with an assumed $\chi=5$
scaling of the hydrodynamic timescale. Results for the Ni isotopes are very similar to the
default ($\chi=1$) scaling case (Figure \ref{CasA_npf}), but due to the longer time spent
at high temperature, the $\alpha$-particle abundance at the onset of $^{40}$Ca assembly
was much lower ($\sim 40$\%) than in the default scaling case, which facilitated a much lower
final value for $P_{44}$ in all cases studied.
}
\label{CasAnpf5}
\end{center}
\end{figure}

\clearpage

\end{document}